\documentclass[11pt,twoside,a4paper,cmspaper,final,collab]{cms-tdr}
\def\svnVersion{ea37516}\def\svnDate{2026/09/29}\def\cmsCernNoTag{CERN-EP-2026-115}\def\cmsCernDate{\today}\def\cmsMessage{Published in the Journal of High Energy Physics as \href{http://dx.doi.org/10.1007/JHEP09(2026)230}{\doi{10.1007/JHEP09(2026)230}.}}
\begin{document}\cmsNoteHeader{B2G-24-020}

\newcommand{\ST}{\ensuremath{S_\text{T}}\xspace}
\newcommand{\mtH}{\ensuremath{m_{\PQt\PH}}\xspace}
\newcommand{\mll}{\ensuremath{m_{\Pell\Pell}}\xspace}
\newcommand{\mo}{\ensuremath{m_{0}}\xspace}
\newcommand{\nwa}{\ensuremath{\Gamma_{\PQT}/{M}_{\PQT}}\xspace}
\newcommand{\pb}{\unit{pb}}
\newcommand{\MT}{\ensuremath{M_{\PQT}}\xspace}

\providecommand{\PQqpr}{\HepParticle{\PQq}{}{\prime}\xspace}
\providecommand{\Pj}{\HepParticle{j}{}{}\xspace}
\title{Search for single vector-like quark production in opposite-sign dilepton final states in proton-proton collisions at \texorpdfstring{$\sqrt{s}=13\TeV$}{sqrt(s) = 13 TeV}}

\author[cern]{A. Cern Person}
\date{\today}

\abstract{A search is presented for single production of a vector-like top quark \PQT, decaying into the standard model top quark and Higgs boson, in a final state including two opposite-sign leptons (electrons or muons), jets, and missing transverse momentum. The data were recorded by the CMS experiment in proton-proton collisions at a center-of-mass energy of 13\TeV at the CERN LHC in the years 2016--2018, and corresponding to an integrated luminosity of up to 138\fbinv. No excess in data over the background expectations is observed. Upper limits at 95\% confidence level on the product of the \PQT production cross section and its decay branching fraction to tH are set. The observed limits are 2.0\pb at a \PQT mass $\MT=600\GeV$ and 0.3\pb at $\MT=1200\GeV$, with a minimum of 0.1\pb at $\MT=1000\GeV$. This is the first search for single vector-like \PQT quark production, followed by a decay to a top quark and the Higgs boson, in opposite-sign dilepton final states.}

\hypersetup{%
pdfauthor={CMS Collaboration},%
pdftitle={Search for single vector-like quark production in opposite-sign dilepton final states in proton-proton collisions at sqrt s = 13 TeV},%
pdfsubject={CMS},%
pdfkeywords={CMS, VLQ, OS}}

\maketitle

\section{Introduction}
\label{sec:introduction}

Vector-like quarks (VLQs)~\cite{Theory,Lenz:2013iha} are quarks for which the left- and right-handed chiral components transform in the same way under the standard model (SM) electroweak group. They are considered as a possible scenario for a fourth generation of quarks, as their gauge-invariant mass terms can be included in the SM Lagrangian without the interaction with the  SM Higgs boson (\PH). They can therefore be heavy states escaping the present constraints from Higgs boson coupling measurements. They are particularly attractive for solving the hierarchy problem and explaining~\cite{Branco:2021vhs, Belfatto:2023tbv} the hints of deviations from the Cabibbo--Kobayashi--Maskawa matrix unitarity indicated by current observations~\cite{Pocanic:2003pf,Czarnecki:2019iwz}. In models of physics beyond the SM, VLQs can appear as two singlets, three doublets, or two triplets~\cite{Aguilar-Saavedra:2013qpa}. They can be produced singly or in pairs in proton-proton (pp) collisions at the CERN LHC, and have been extensively searched for by the ATLAS and CMS Collaborations, as summarized in Refs.~\cite{ATLAS:2024xdc,CMS:2024bni}.

\begin{figure}[b]
    \centering
    \includegraphics[width=0.49\textwidth]{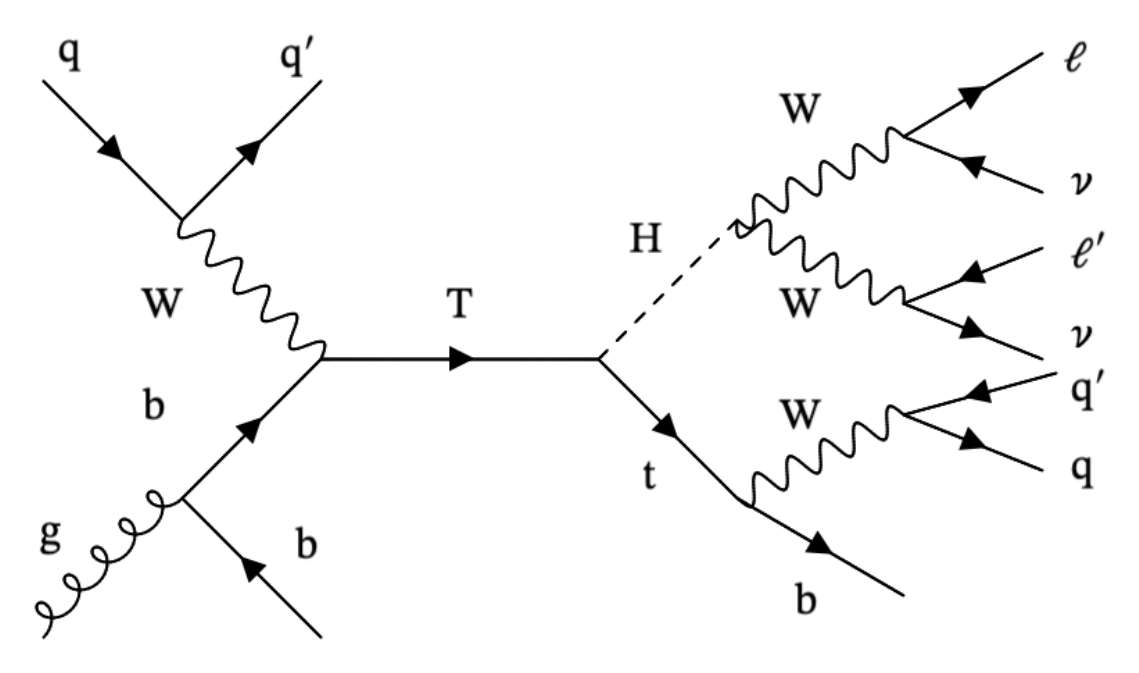}
    \hfill
    \caption{Leading-order Feynman diagram for the single production of a vector-like \PQT quark via a \PW boson and further decaying into a dilepton final state.}
\label{fig:diagram}
\end{figure}

The electroweak single \PQT VLQ production ($\Pp \Pp \to \PQT \PQb \PQq $, as shown in Fig.~\ref{fig:diagram}) proceeds through a $t$-channel exchange of a \PW boson, and its cross section depends on the coupling between the VLQ, \PW boson, and a \PQb quark. Under certain assumptions (as discussed in Section 5.2 of Ref.~\cite{CMS:2024bni}), the width-to-mass ratio, \nwa, can be used to implicitly fix this coupling. The VLQ denoted by \PQT has a charge of $+2/3$, and can decay as $\PQT \to \PQb \PW$, $\PQT \to \PQt \PZ$, or $\PQT \to \PQt \PH$.
The branching fractions for these decay modes are assumed to be 50\%, 25\%, and 25\%, respectively, as motivated by the model in Ref.~\cite{Aguilar-Saavedra:2009xmz}. In this paper, we focus on the $\PQT \to \PQt \PH$ process.
The \PQb quark accompanying the VLQ is typically produced with much smaller transverse momentum (\pt) than the VLQ decay products, and is therefore not considered in the selection.

The three decay modes mentioned above have been sought in previous LHC analyses, as noted in Refs.~\cite{ATLAS:2024xdc,CMS:2024bni}. These decay modes lead to a number of different final states. Specifically, for $\PQT \to \PQt \PH$, previous CMS analyses have considered final states where the top quark decays hadronically and the Higgs boson decays into two \PQb quarks or two photons~\cite{CMS:2019afi, CMS:2023agg, CMS:2024qdd}. As a result of the dependence of the single VLQ production cross section on the couplings to SM quarks and electroweak bosons, a model-independent limit can be set only on the product of the cross section and branching fraction, as a function of the different VLQ masses. Various \PQT decay width hypotheses were taken into consideration, including the narrow-width approximation (with \nwa typically around 5\%).

The single \PQT search presented here covers the \PQT mass range between 600 and 1200\GeV, and opposite-sign (OS) dilepton final states are probed for the first time in single VLQ searches at the LHC. The \PQT resonance is assumed to have a narrow width, and its decay products are considered to be resolved in this \PQT mass range. The analysis targets OS dilepton ($\Pellp\Pellmpr$, where $\Pell,\Pellpr = \Pe$ or \PGm) final states of the process where the top quark decays fully hadronically and the Higgs boson decays primarily into a pair of \PW bosons, each decaying leptonically, resulting in an OS dilepton pair in the final state (including the contribution from the \PW bosons decaying into \Pgt leptons, which further decay leptonically). In addition, we consider smaller contributions to the OS dilepton final state originating from the Higgs bosons decaying to \Pgt lepton pairs and pairs of \PZ bosons, as detailed in Section~\ref{Sec:SignalModel}. These processes result in three different topologies: the dielectron, dimuon, and electron-muon final states, accompanied by at least three jets, of which at least one originates from a \PQb quark, and missing transverse momentum \ptmiss. 

The event selection is optimized to suppress the main backgrounds, which are mainly low-mass Drell--Yan (DY) production for the $\Pe \Pe$ and $\PGm \PGm$ channels, and top quark pair production for all three channels. The reconstructed mass of the \PQT candidate is used as the discriminating variable in the final likelihood fit. The background is determined from the observed data, with a shape parameterization motivated by Monte Carlo (MC) simulations.

The paper is structured as follows: Section~\ref{sec:CMSdDetector} describes the CMS detector. Event reconstruction is discussed in Section~\ref{sec:CMSReco}. The data and the simulated samples used are presented in Section~\ref{sec:DataMC}. Section~\ref{sec:Selection} details the event selection strategy, including trigger usage, preselection, main selections, and signal reconstruction. The signal modeling and background estimation are introduced in Sections~\ref {Sec:SignalModel} and~\ref{Sec:Background}, respectively. The method of signal extraction and the study of systematic uncertainties are described in Section~\ref{Sec:Systematics}. Finally, the results of the search are presented in Section~\ref{Sec:results}, followed by a summary in Section~\ref{sec:Summary}. Tabulated results can be found in the HEPData record for this analysis~\cite{HEPData}. 

\section{The CMS detector}
\label{sec:CMSdDetector}
The CMS apparatus is a multipurpose detector designed around a superconducting solenoid of 6\unit{m} internal diameter, providing a magnetic field of 3.8\unit{T}.  Within the solenoid volume are a silicon pixel and strip tracker, a lead tungstate crystal electromagnetic calorimeter (ECAL), and a brass and scintillator hadron calorimeter (HCAL), each composed of a barrel and two endcap sections. Forward calorimeters extend the pseudorapidity coverage to $\abs{\eta}<5.2$. Muons are reconstructed using gas-ionization detectors interleaved with the layers of the steel flux-return yoke outside the solenoid. More detailed descriptions of the CMS detector, together with a definition of the coordinate system used and the relevant kinematic variables, can be found in Ref.~\cite{CMS:2008xjf,CMS:2023gfb}.

The CMS two-tiered trigger system is composed of the level-1 (L1) trigger and the high-level trigger (HLT)~\cite{CMS:2016ngn}. The L1 trigger, based on a set of hardware processors, uses information collected by the muon detectors and the calorimeters to select events at a rate of around 100~kHz within a fixed latency of 4\mus~\cite{CMS:2020cmk}. The HLT consists of a large array of processors, running a version of the full event reconstruction software optimized for fast processing, and reduces the stored event rate to a few kHz~\cite{CMS:2016ngn,CMS:2024aqx}.

\section{Event reconstruction}
\label{sec:CMSReco}
The particle-flow (PF)~\cite{CMS:2017yfk} algorithm aims to reconstruct and identify individual particles  (electrons, muons, photons, and charged and neutral hadrons) in an event, referred to as PF candidates, using an optimized combination of information from the various CMS subdetectors. This analysis is based on many types of reconstructed objects, including electrons, muons, jets, and \ptmiss. The primary pp interaction vertex is taken to be the vertex corresponding to the most energetic interaction in the bunch crossing, evaluated using tracking information alone, as described in Ref.~\cite{CMS:2014pgm}.

Electrons are reconstructed using the momentum and direction at the primary interaction vertex as determined by the tracker, the energy of the corresponding ECAL cluster, and the energy sum of all bremsstrahlung photons spatially compatible with those having originated  from the electron track.
Muons are identified as tracks in the central tracker consistent with either a track or several hits in the muon system, and associated with calorimeter deposits compatible with the muon hypothesis. The energy of a muon is obtained from the curvature of the corresponding track. The energy of a charged hadron is determined from a combination of the momentum measured in the tracker and the matching ECAL and HCAL energy deposits, corrected for the response function of the calorimeters to hadronic showers. Finally, the energy of neutral hadrons is obtained from the corresponding corrected ECAL and HCAL energies.

The hadronic jets are found from clusters of these reconstructed particles using the infrared and collinear safe anti-\kt algorithm~\cite{Cacciari:2008gp, Cacciari:2011ma}. The analysis utilizes a distance parameter of 0.4.
Jet momentum is determined as the vectorial sum of all particle momenta in the jet, and is found from simulation to be, on average, within 5--10\% of the true momentum over the entire \pt spectrum and detector acceptance. Additional $\Pp\Pp$ interactions within the same or nearby bunch crossings (pileup) can contribute additional tracks and calorimetric energy depositions to the jet momentum. To mitigate this effect, charged particles identified to be originating from pileup vertices are discarded and an offset correction is applied to correct for remaining contributions. Jet energy corrections are derived from simulation to bring the measured response of jets to that of particle-level jets on average. In situ measurements of the momentum balance in dijet, $\text{photon} + \text{jet}$, $\PZ + \text{jet}$, and multijet events are used to account for any residual differences in the jet energy scale (JES) between data and simulation~\cite{CMS:2016lmd}. The jet energy resolution (JER) amounts typically to 15--20\% at 30\GeV, and 10\% at 100\GeV~\cite{CMS:2016lmd}.
Additional selection criteria are applied to each jet to remove jets potentially dominated by anomalous contributions from various subdetector components or reconstruction failures~\cite{CMS:2020ebo}.

The missing transverse momentum vector \ptvecmiss is computed as the opposite of the vector sum of the transverse momenta of all the PF candidates in an event, and its magnitude is denoted as \ptmiss~\cite{CMS:2019ctu}. The \ptvecmiss is modified to account for corrections to the energy scale of the reconstructed jets in the event.

The identification of jets resulting from the hadronization of \PQb quarks is performed using the \textsc{DeepJet} algorithm~\cite{Bols:2020bkb,CMS-DP-2023-005,CMS:2017wtu}, which provides a unity-normalized score for different jet flavors. A specific value requirement for a jet on the score of the algorithm is referred to as \PQb tagging, and positively identified jets are referred to as \PQb-tagged jets or \PQb jets.
In this analysis, the medium working point of the \textsc{DeepJet} algorithm is chosen. This working point corresponds to an expected \PQb jet identification efficiency of ${\approx}80$\% for an expected misidentification rate for jets originating from light-flavor quarks and gluons (\PQc quarks) of 1\% (15)\%~\cite{CMS-DP-2018-058,CMS-DP-2023-005}.

\section{Data and simulated samples}
\label{sec:DataMC}
The analysis uses pp collision data collected by the CMS experiment at $\sqrt{s}=13\TeV$ in the years 2016, 2017, and 2018, corresponding to an integrated luminosity of up to 138\fbinv. Four periods of data-taking (two in 2016, and one each in 2017 and 2018) were analyzed and simulated separately to match the appropriate detector operation conditions and calibrations. For the $\Pe \Pe$ final state, the integrated luminosity was 133\fbinv because the single-electron trigger for this channel was disabled for a short period of time in 2017.

The signal process $\Pp \Pp \to \PQT \PQb \PQq$ is generated with the \MGvATNLO generator version 2.6.5~\cite{Alwall:2014hca} at leading order (LO).
For simplicity, we set the \PQT width to 10\GeV for all \PQT mass hypotheses, which is negligible compared to the experimental resolution, and we set the masses of Higgs boson and top quark to 125 and 172.5\GeV, respectively. The signal samples are generated for seven mass points ranging from 600 to 1200\GeV, in steps of 100\GeV. The Higgs boson is forced to decay to a \PW boson pair, a \PZ boson pair, or a \PGt lepton pair, since these are the only decay modes significantly contributing to the OS dilepton final state, while the top quark is allowed to decay to all possible channels.

Regarding background events, the top quark pair (\ttbar) production process provides major contributions to all three channels ($\Pe \Pe$, $\PGm \PGm$, $\Pe \PGm$), while DY production mainly contributes to the $\Pe \Pe$ and $\PGm \PGm$ channels. The \ttbar MC sample is generated at next-to-leading order (NLO) with \POWHEG version 2.0~\cite{Alioli:2010xd, Frixione:2002ik}. As the production of simulated samples at NLO requires significantly more computing resources than the production of LO samples with equivalent statistical power, the DY MC samples are generated at LO by \MGvATNLO for better statistical precision. A correction to the DY sample calculated using a $K$-factor is introduced to the DY sample to approximate the NLO-level description. Other minor backgrounds are
single top quark  ($\PQt\PW$), diboson ($\PV\PV$, with $\PV=\PW \text{ or }\PZ$), and $\ttbar \PV$ production. 
 These simulated samples are used for optimizing the selection strategy, but not for background estimation in the signal extraction, which is performed using the observed data.

For all processes, the hard interactions are simulated with the NNPDF 3.1 set of parton distribution functions (PDFs) at next-to-NLO (NNLO) accuracy~\cite{NNPDF:2017mvq}, and the parton shower is  simulated by \PYTHIA, version 8.240~\cite{Sjostrand:2014zea}. Pileup events are generated with \PYTHIA and are added according to the expected pileup profile in the observed data. All generated events are passed through a \GEANTfour-based~\cite{GEANT4:2002zbu} simulation of the CMS detector and reconstructed using the same version of the CMS event reconstruction software used for collision data.

\section{Event selection}
\label{sec:Selection}
This analysis targets the OS dilepton final state of the following process and decay chain, where the top quark decays fully hadronically:  $\Pp \Pp \to \PQT \PQb \PQq$; $\PQT \to \PQt \PH$; $\PQt \to \PQb \PW \to \PQb \PQq \PQqpr$.
The primary signal is provided by this process, with the Higgs boson decay mode $\PH\to \PW \PW \to \Pell\Pellpr \PGn \PGn$. The final state, therefore, includes two charged leptons, three jets, with at least one of them a \PQb-tagged jet, and missing transverse momentum. As detailed in Section~\ref{Sec:SignalModel}, the decay modes $\PH\to \PGt\PGt$ and $\PH\to \PZ \PZ$ give further, albeit smaller, contributions to the OS dilepton final state, and are both also regarded as signal in this analysis. The contributions of other Higgs boson decay modes are found to be negligible.

All events in the analysis must pass the HLT selection.
The online selection for the $\Pe\Pe$ category is based on a combination of a low-\pt single-electron trigger with an isolation requirement, with a minimum \pt threshold of 27--32\GeV, depending on the data-taking year, and a high-\pt single-electron trigger without the isolation requirement, with a minimum \pt threshold of  115\GeV. For the $\Pgm\Pgm$ and $\Pe\Pgm$ channels, an analogous combination of single-muon triggers is used, with minimum \pt thresholds of 24--27\GeV, depending on the data-taking year, and 50\GeV for the respective single-muon triggers. After the online selections, the preselection selects the topologies for signal reconstruction, and the subsequent main selections are optimized to suppress background events.

\subsection{Preselection}
The preselection requires exactly two OS leptons and at least three jets, including at least one \PQb-tagged jet in the event.
Electron candidates must have the $\pt > 35\GeV$ and $\abs{\eta} < 2.5$, have an impact parameter significance compatible with the primary vertex, and must satisfy tight electron identification requirements~\cite{CMS:2020uim}. Electron candidates are vetoed if they lie within the transition region between the barrel and endcap of the ECAL, defined by $1.444<\abs{\eta}<1.566$.
Muon candidates must have $\pt>30\GeV$, motivated by the online selection, and they must satisfy tight identification and isolation requirements~\cite{CMS:2018rym}. All muons are required to pass the $\abs{\eta} < 2.4$ selection because of the acceptance of the muon detector.
The jets must satisfy $\pt>30\GeV$, $\abs{\eta} < 2.5$, and tight identification requirements~\cite{CMSjetID}. The jets with $\pt < 50\GeV$ need to pass loose selections to reduce the impact of pileup. Furthermore, jets are removed if they overlap with an isolated lepton, which is the case if their axis lies within a cone of $\Delta R(\Pj, \Pell ) = \sqrt{(\Delta \eta (\Pj, \Pell ))^2 + (\Delta \phi (\Pj, \Pell ))^2}< 0.4$ around the latter. To remove contributions from meson resonances decaying to lepton pairs in the low-mass region, the invariant mass of the lepton pair must be larger than 12\GeV.

\subsection{Signal reconstruction}
The heavy resonance probed in this analysis, \PQT, decays into a top quark and a Higgs boson and the reconstructed mass of the \PQT candidate, \mtH,  is taken as the main discriminating variable. Since the top quark is selected in its fully hadronic decay, the reconstructed masses of the \PW and top quark candidates can be used to select the optimal three-jet combination.
In a first pass, of all jets passing the preselection, the pair with the invariant mass closest to the world-average \PW boson mass~\cite{ParticleDataGroup:2024cfk} is chosen to reconstruct the \PW boson candidate. In a second pass, among all remaining jets, the ones that are \PQb tagged are paired with the \PW candidate, and the combination with the invariant mass closest to the world-average  top quark  mass~\cite{ParticleDataGroup:2024cfk} is chosen to reconstruct the top quark candidate.

In the Higgs boson decay mode $\PH\to \PW \PW \to \Pell \Pellpr \PGn \PGn$, the final state cannot be fully reconstructed because of the two neutrinos. We address this by a partial reconstruction, exploiting the fact that the kinematic properties of the neutrino pair are similar to those of the lepton pair~\cite{CMS:2021roc}. As the contribution from \Pgt induced electrons and muons and from other considered Higgs boson decays is small, they are reconstructed in the same way despite a somewhat different final-state topology.

\begin{figure}[htbp]
    \centering
    \includegraphics[width=0.45\textwidth]{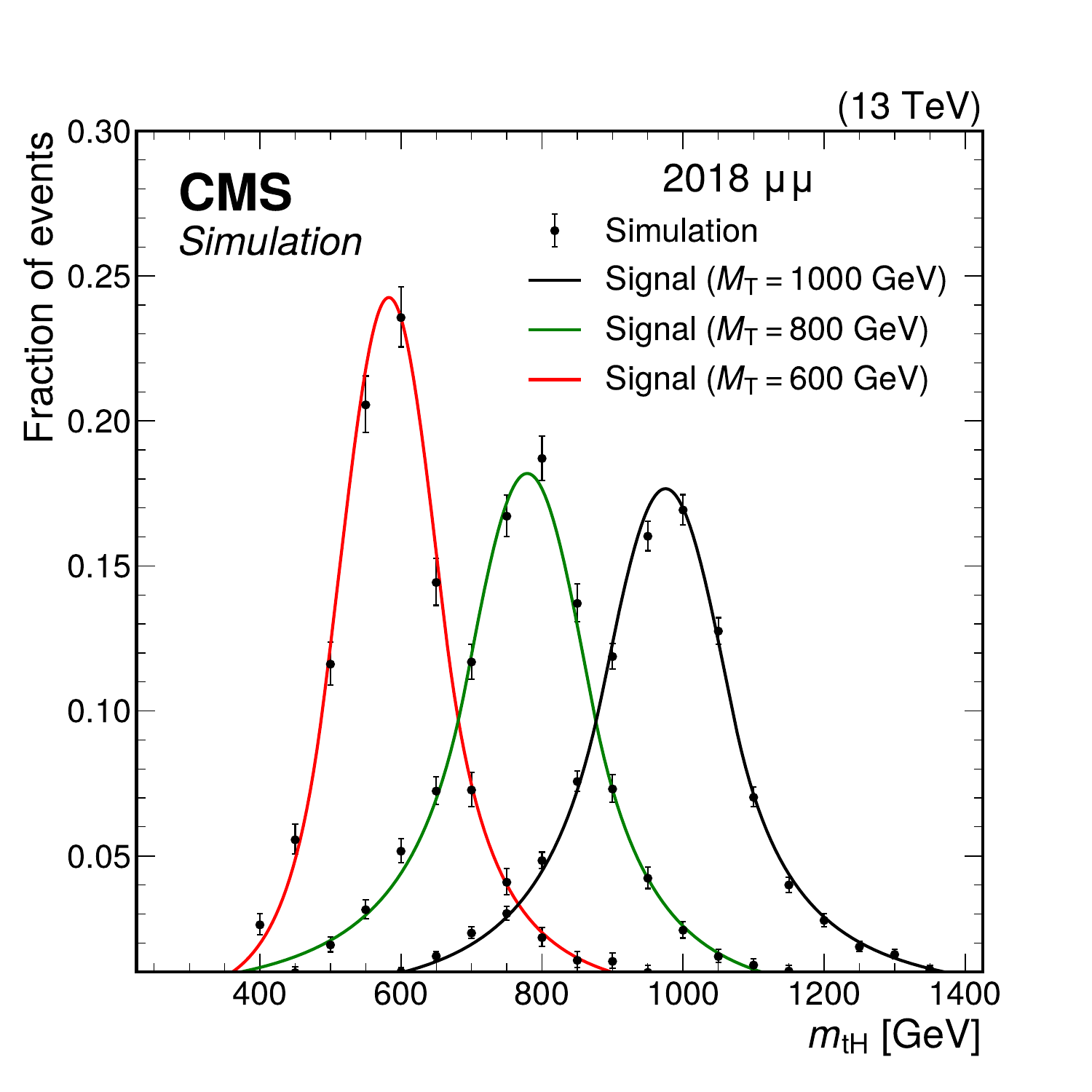}
    \hfill
    \caption{Distributions of the reconstructed invariant mass in simulated signal events for \PQT masses of 600, 800, and 1000\GeV, for the $\Pgm\Pgm$ channel and the 2018 data-taking conditions, as a representative example. Signal histograms are normalized to unity. Each distribution is shown with its nominal normalization, and can be fitted with a double-sided Crystal Ball function~\cite{Oreglia:1980cs,Gaiser:1982yw}, shown as red, green, and black solid lines.}
\label{fig:Signal}
\end{figure}

For the \PQT mass values studied in this analysis, the neutrinos can be considered approximately collinear with the lepton pair stemming from the same Higgs boson decay. Thus, one can approximate the polar angle to the beam direction as $ \theta_{\Pell \Pellpr} = \theta_{\PGn  \PGn}$. Based on this assumption, the $p_\text{z}$ of the neutrino pair is calculated according to
$p_\text{z}^{\PGn  \PGn} = \ptmiss / \tan \theta_{\PGn  \PGn}=\ptmiss / \tan \theta_{\Pell \Pellpr}$.
The only missing kinematic variable is the invariant mass of the neutrino pair. It is approximated by the mean value of its distribution obtained from the simulated signal events, which is found to be $\langle m_{\PGn  \PGn} \rangle = 30\GeV$. This number is used for the reconstruction of all Higgs boson candidates. Thanks to the reconstruction method based on the neutrino kinematic assumption, the \PQT mass variable peaks near the nominal mass, as shown in Fig.~\ref{fig:Signal}, and is well described by a double-sided Crystal Ball function~\cite{Oreglia:1980cs,Gaiser:1982yw}.

\subsection{Main selections}
\label{Sec:Selection}
The main selection criteria detailed in the following are designed and optimized by utilizing Punzi significance~\cite{Punzi:2003bu} based on the differences in kinematics between the signal and background processes, using the $\MT=700\GeV$ signal as a benchmark point. The dominant remaining background processes arise from DY and \ttbar production. As the invariant mass of the OS lepton pair, \mll, from the signal process is smaller than that from \ttbar production, the selection requires  $\mll<60\GeV$. Moreover, the invariant mass distribution of the lepton pairs from the DY process peaks near the \PZ boson mass. Both background sources are therefore significantly suppressed by this selection.

In the signal process, the lepton and the \PQb quark originate from different mother particles, whereas in the \ttbar process, they emerge from the same top quark decay. To further reduce the \ttbar background, we require that $\Delta R_\text{min}(\PQb^\PQt, \Pell)>2$, where $\Delta R({\PQb}^{\PQt}, \Pell)$ is the spatial separation between the selected \PQb jet (${\PQb}^{{\PQt}}$) used for the top reconstruction and the closer of the two leptons.

In the process $\PQT \to \PQt \PH$, the top quark's hadronic decay products are naturally collimated, while there are no hadronically decaying top quarks in any of the main background processes. We thus require $\Delta R({\PQb}^{\PQt}, \PW^{\PQt})<2.5$, where $\PW^{\PQt}$ is the \PW boson candidate in the reconstruction of the top quark candidate. This selection suppresses further both the \ttbar and DY background processes.

A strong discriminating variable is the quantity \ST, which is defined as the scalar sum of the \pt values of the Higgs boson and the top quark candidates. Since the \PQT particle is a heavy resonance, its decay products tend to have much higher \pt than those from the background processes. However, the reconstructed mass of the \PQT candidate is correlated with the \pt of its decay products, and thus with \ST. A fixed threshold for \ST would thus modify the otherwise monotonically falling shape of the background such that a peak may emerge in the signal region, which would affect the signal extraction. To mitigate this issue, the \ST threshold, ${\ST}^{\text{min}}$, is chosen to be a function of \mtH instead of a fixed value. Based on simulation, this function is chosen such that the background selection efficiency becomes independent of the reconstructed \mtH for the background candidates passing the signal selection, while maintaining high sensitivity to the signal. The \ST threshold corresponding to a given signal efficiency as a function of the reconstructed \mtH is fitted with a polynomial function. 

The resulting selection is thus $\ST > {\ST}^{\text{min}}(\mtH)$, where
\begin{equation}
{\ST}^{\text{min}}(\mtH) = 72.9\GeV+0.18\mtH+6.07 \times 10^{-4}\GeV^{-1}m^2_{\PQt\PH}-3.47 \times 10^{-7}\GeV^{-2}m^3_{\PQt\PH} .
\end{equation}

The above-mentioned selection strongly reduces the background from DY and \ttbar production. The signal efficiency for a \PQT with a mass of 900\GeV is about 33\% for the $\Pe\Pe$ channel and 44\% for the $\Pgm\Pgm$ and $\Pe\Pgm$ channels.

\section{Signal model}
\label{Sec:SignalModel}
In addition to the decay mode $\PH\to \PW \PW$, which clearly dominates this search, the $\PH \to \PGt \PGt$ and $\PH\to \PZ \PZ$ decay modes contribute to the signal. A detailed breakdown of the decay chains and corresponding contributions after the preselection and the $\mll<60\GeV$ requirement is given in Table~\ref{Tab:SignalChannels}. The $\PH\to \PZ \PZ$ decay mode contributes noticeably only to the $\Pe\Pe$ and $\PGm \PGm$ channels. While signatures with semileptonic top quark decays are included in the signal generation, they are removed by these selections.

\begin{table}[htbp]
\begin{center}
  \topcaption{Higgs boson decay modes contributing to the OS dilepton signal.
The numbers in the two rightmost columns show the relative contributions of the $\PH\to \PW \PW$, $\PH\to \PGt\PGt$, and $\PH\to \PZ \PZ$ decay modes before the $\Delta R_\text{min}({\PQb}^{\PQt}, \Pell)$ selection discussed in Section~\ref{Sec:Selection} is applied.}
\begin{tabular}{lllcc}
\hline
H decay mode & & Final state & \multicolumn{2}{c}{Relative contribution} \\
          & &  &$\Pe\Pe / \PGm \PGm$&$\Pe \PGm$\\
\hline
$\PH\to \PW \PW$ & $\to \Pell \Pellpr \Pgn \Pgn$& & 80\% & 90\%\\
                    & $\to \PGt \Pgn  \Pell \Pgn$
                    &  $\Pell \Pellpr + \ptmiss$ & &  \\
                    & $\to \PGt \Pgn \PGt \Pgn$
                    &  & &  \\
\\
$\PH\to \PGt\PGt$ &  & $\Pell \Pellpr + \ptmiss$ & 5\% & 10\% \\
\\
$\PH\to \PZ \PZ$ & $\to \Pell \Pell \Pgn \Pgn$& $\Pell \Pell + \ptmiss$ & 15\% & ${\approx}0\%$\\
                    & $\to \PGt  \PGt \Pgn \Pgn$ & $\Pell \Pellpr + \ptmiss$ & &  \\
                    & $\to \Pell \Pell \PQq \PQqpr$ & $\Pell \Pell \PQq \PQqpr$ & & \\
                    & $\to \PGt  \PGt \PQq \PQqpr$ & $\Pell \Pellpr \PQq \PQqpr + \ptmiss$ & &  \\
\hline
\end{tabular}
\label{Tab:SignalChannels}
\end{center}

\end{table}

Models of the signal are obtained by fitting the reconstructed tH invariant mass distributions from the simulated signal events for each mass point, dilepton channel and data-taking period with a double-sided Crystal Ball function. Examples of the signal models for three mass points are shown in Fig.~\ref{fig:Signal}. The signal shapes, including the tails, are found to be well modeled by the fit functions.

\section{Background estimation}
\label{Sec:Background}
\begin{figure}
    \centering
    \includegraphics[width=0.45\textwidth]{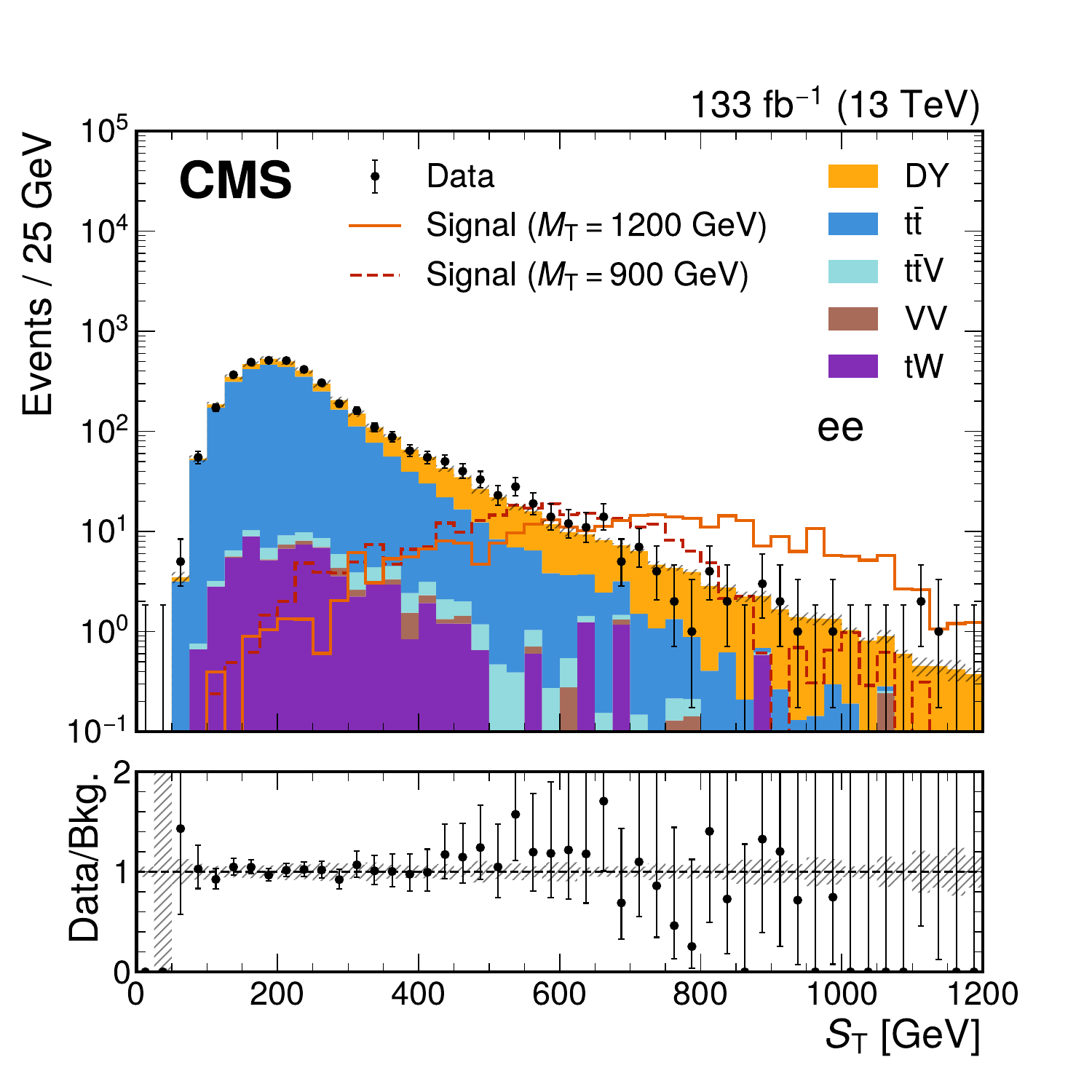}
    \includegraphics[width=0.45\textwidth]{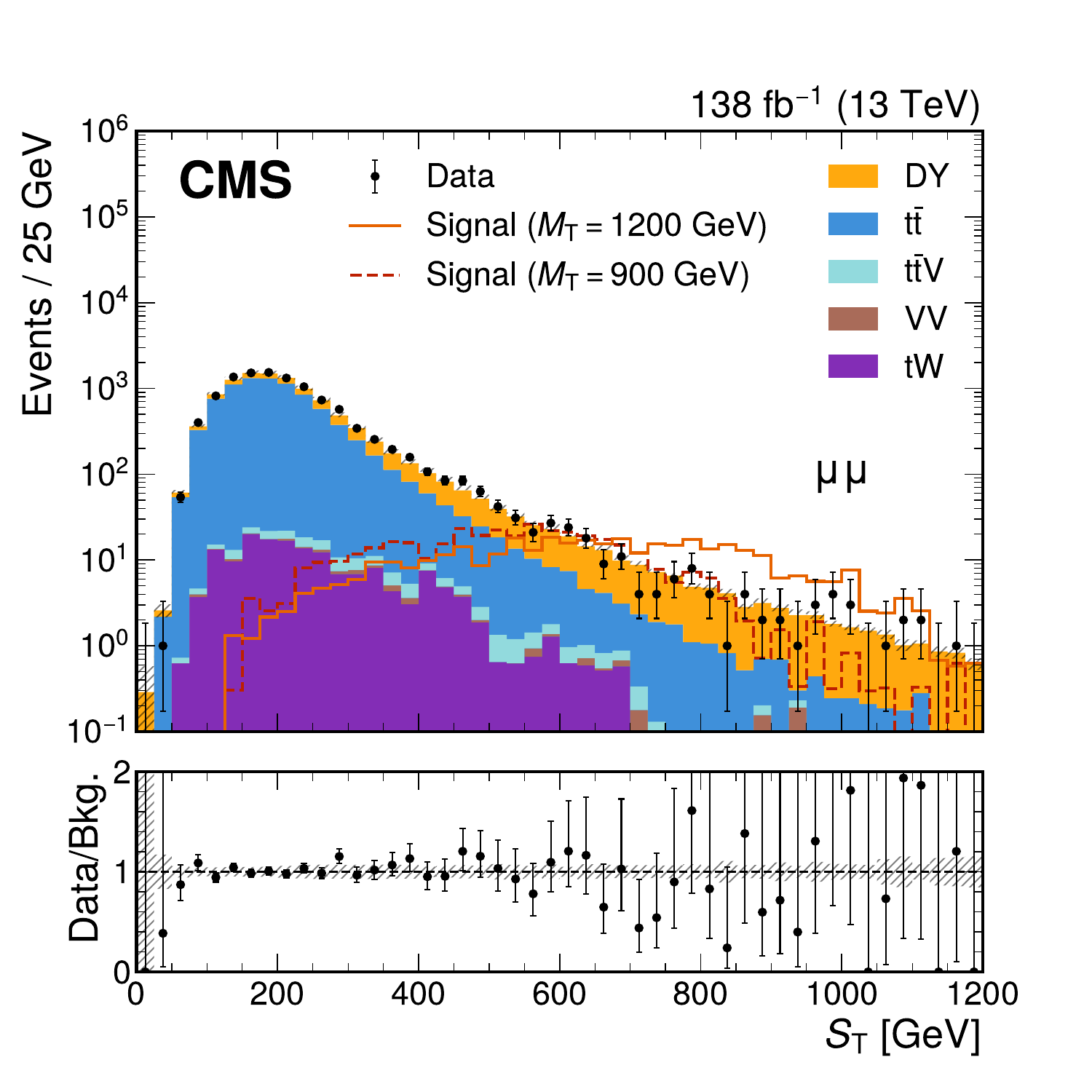}
    \includegraphics[width=0.45\textwidth]{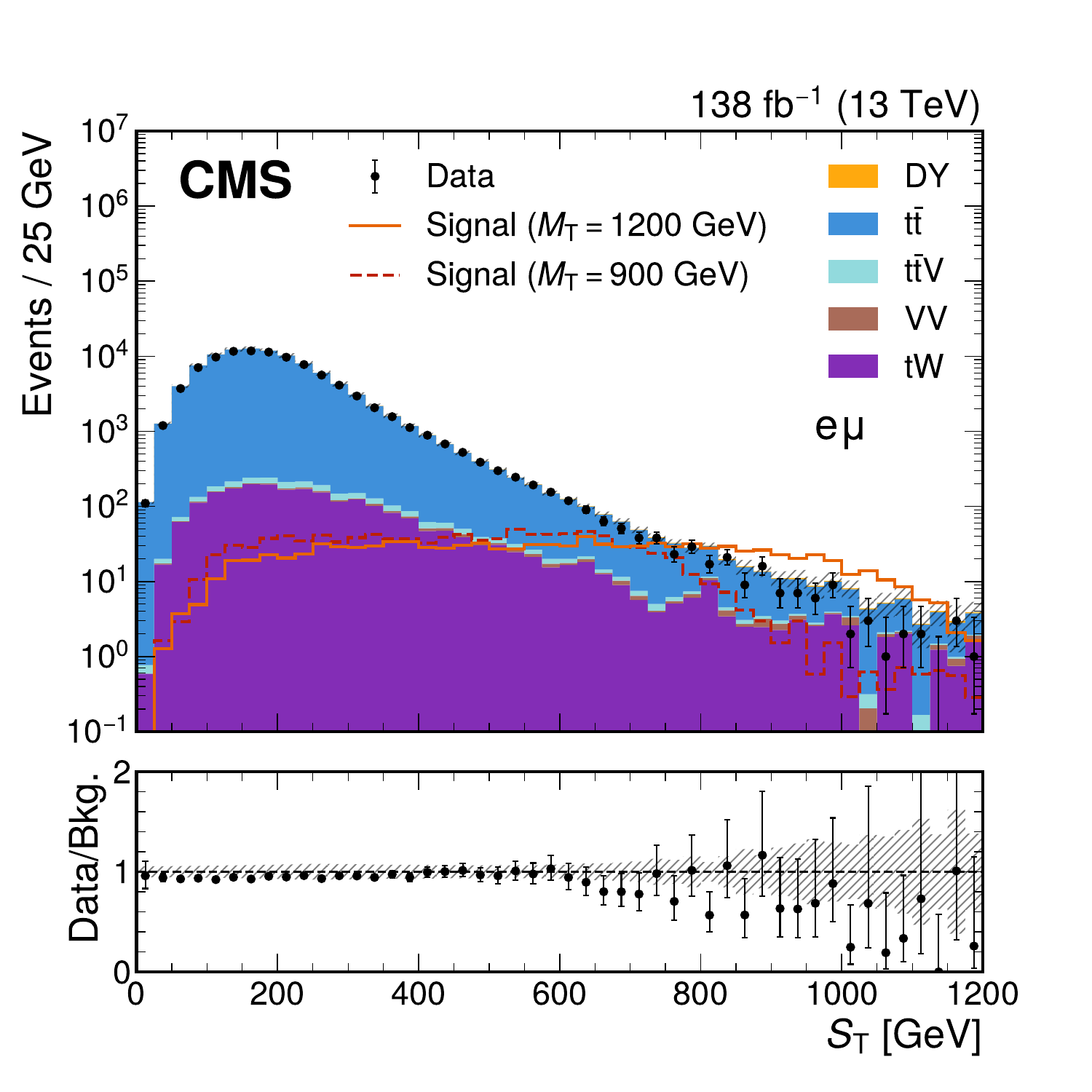}
    \hfill
    \caption{Comparison of the observed data and simulated background distributions of the variable \ST for the 2016--2018 data, with two representative signal benchmark distributions overlaid. Shown are the distributions for the $\Pe\Pe$ (upper left) and  $\PGm\PGm$  (upper right) channels after the $\mll<60\GeV$ selection, and for the $\Pe \PGm$ channel after the preselection (lower). The respective lower panels show the ratio of data to simulation (sum of all colored contributions). The shaded bands indicate the total uncertainty in the MC simulation. For all the signal distributions in the plots, the $\sigma(\Pp \Pp\to \PQT) {\mathcal{B}}(\PQT \to \PQt\PH)$ is set to 1\pb.}
\label{fig:dataMC}
\end{figure}

In order not to depend too much on the details of simulation in a highly restricted phase space, we estimate the background in the signal region using the observed data. The simulated background samples are only used for the determination of the initial background shape and the optimization of the selection, in particular regarding the \ST variable. The distributions of the relevant selection variables show good statistical agreement between the observed and simulated data. As an example, Fig.~\ref{fig:dataMC} shows good agreement of the observed distribution of the \ST variable after preselection (and $\mll<60\GeV$ in the $\Pe\Pe$ and $\Pgm\Pgm$ channels) with the simulated distribution, within the uncertainties. Figure~\ref{fig:dataMC} also shows that the DY and \ttbar processes dominate the background for the $\Pe\Pe$ and $\Pgm\Pgm$ channels, while \ttbar production alone is the dominating process in the $\Pe\Pgm$ channel. The contributions from $\PQt\PW$, $\PV\PV$, and $\ttbar\PV$ production processes are negligible, after the preselection.

\begin{figure}
    \centering
    \includegraphics[width=0.45\textwidth]{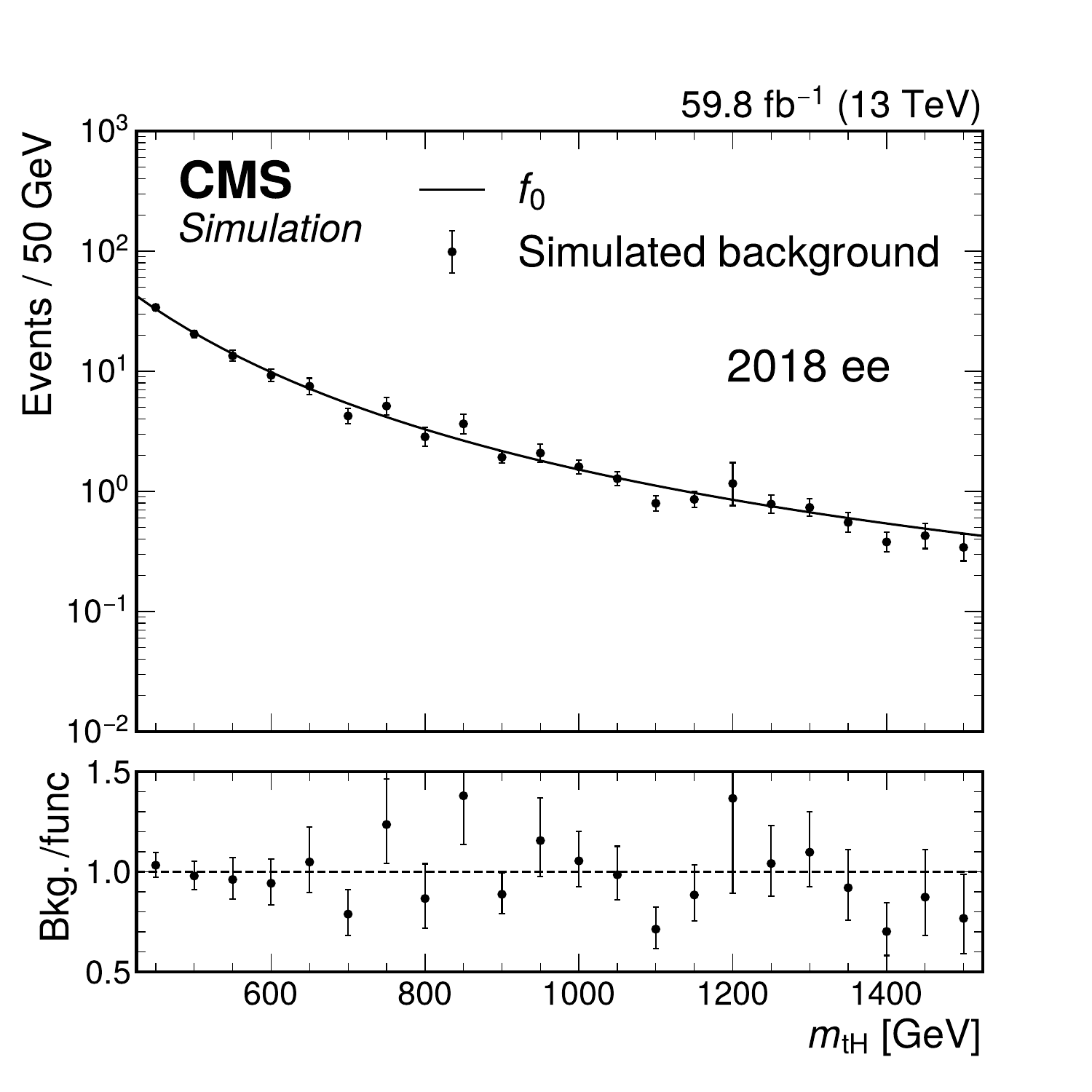}
    \includegraphics[width=0.45\textwidth]{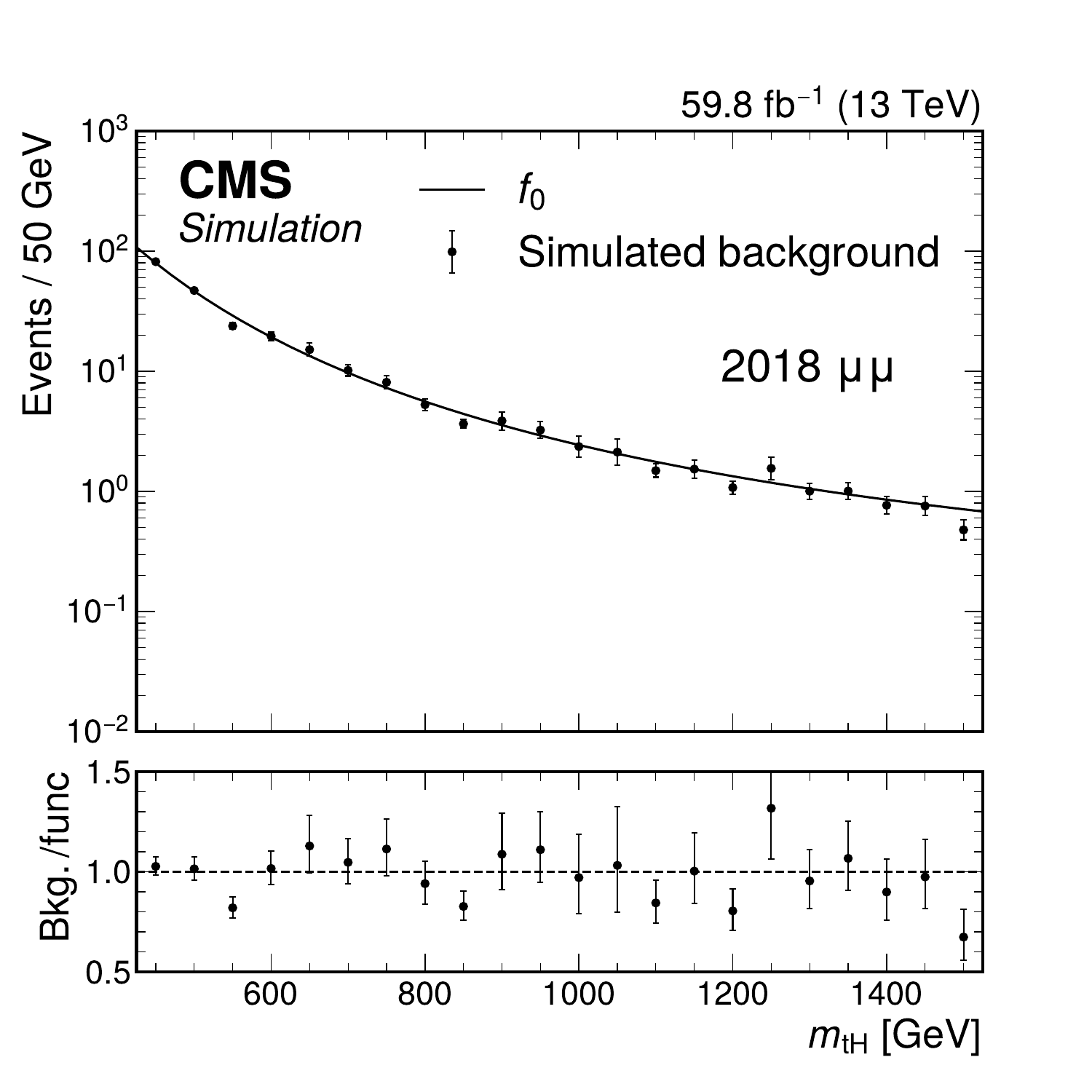}
    \includegraphics[width=0.45\textwidth]{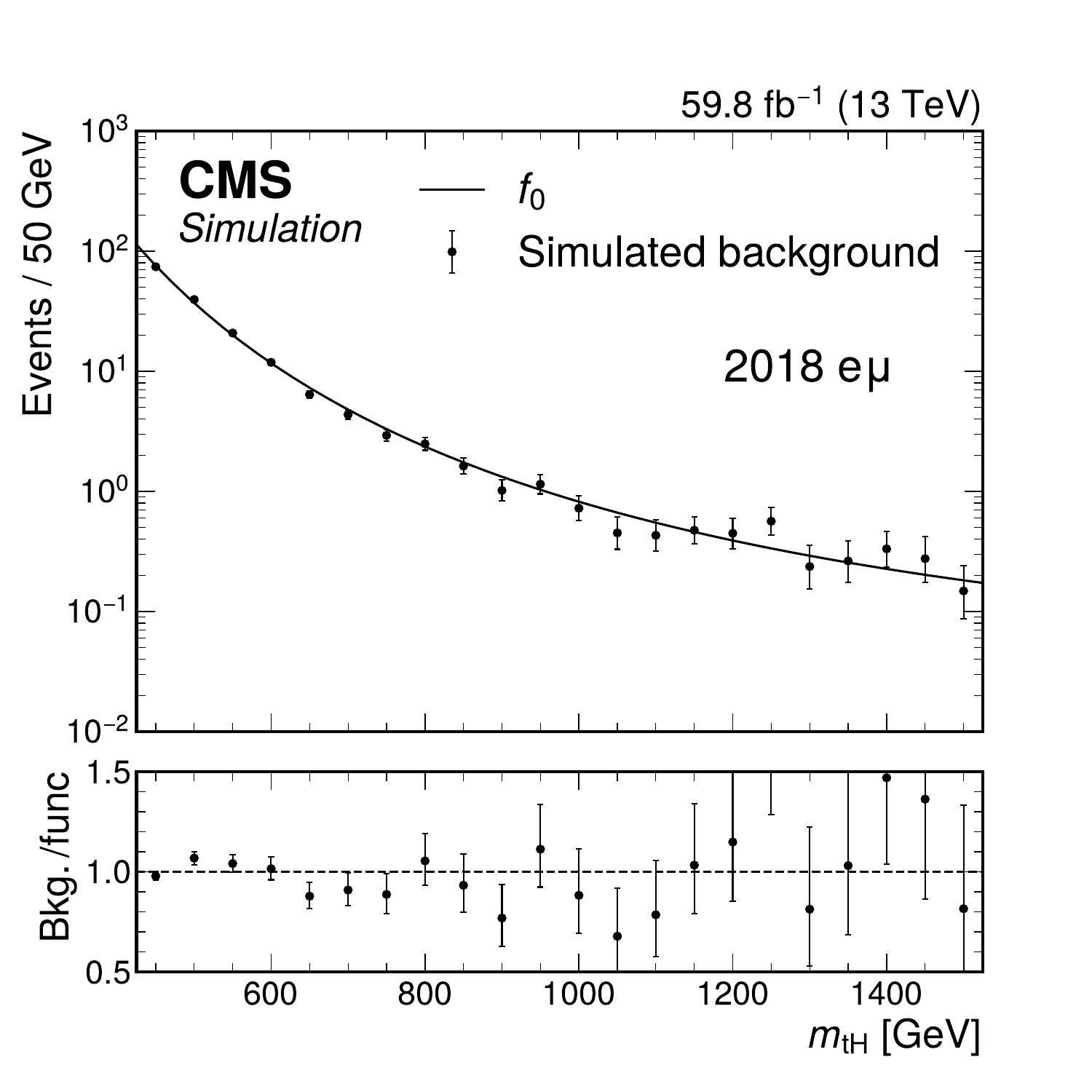}
    \hfill
    \caption{Distribution of the reconstructed mass \mtH in the background simulation, taking the channels $\Pe \Pe$ (upper left), $\PGm \PGm$ (upper right), and $\Pe \PGm$ (lower)
      for the 2018 data, as an example. Superimposed as solid lines are the fits to the background model given by the function $f_0$ in Eq.~(\ref{eq:bkgshape}). The lower panels show the ratio of the MC distribution to the background function fit.}
\label{fig:SigBkg}
\end{figure}

Thanks to the \mtH-dependent threshold of the \ST selection, the latter does not change the shape of the background distribution, which displays a smoothly falling shape, as shown in Fig.~\ref{fig:SigBkg}. This shape is parameterized with an exponential of a second-order polynomial in the logarithm of the invariant mass given by the function $f_0$ in Eq.~(\ref{eq:bkgshape})

\begin{equation}
    f_0(\mtH; \alpha_1, \alpha_2) = \exp{\left[\alpha_1 \Bigl(\ln{\frac{\mtH}{\mo}}\Bigr)+\alpha_2 {\Bigl(\ln{\frac{\mtH}{\mo}}\Bigr)}^2\right]} \label{eq:bkgshape}
\end{equation}

where the value of the constant $\mo=650\GeV$ minimizes the correlation of the parameters. As shown in Fig.~\ref{fig:SigBkg}, this function describes the background shape well in the simulation. The background shapes in all the analysis categories are parameterized separately.

The parameters of the background function are treated as free parameters in the signal extraction fit, and thus the background is estimated directly from observed distributions. As the choice of the functional form may possibly cause a bias and give different results for the signal strength, a systematic uncertainty is associated with the choice of the function. This uncertainty is accounted for with the discrete profiling method~\cite{Dauncey:2014xga}. This method introduces an additional (discrete) nuisance parameter indicating the choice of the background, which is profiled in an analogous way to continuous nuisance parameters. An alternative to the $f_0$ function considered is as follows,

\begin{equation}
f_1(\mtH; \alpha_1, \alpha_2) = \exp{\left[\alpha_1 \Bigl(\frac{\mtH}{\mo}-1 \Bigr)+\alpha_2 {\Bigl(\frac{\mtH}{\mo}-1\Bigr)}^2\right]}.
\end{equation}

The function $f_1$ has been chosen since it represents a different family of functions compared to $f_0$, while still giving an adequate description of the background.

\section{Signal extraction and systematic uncertainties}
\label{Sec:Systematics}
The signal extraction is performed with a binned maximum likelihood fit to the \mtH distribution in a set of categories in the data. The categories correspond to the three lepton flavor combinations in each of the four data-taking periods, yielding a total number of twelve. The fit function is the sum of the signal and background parameterizations, as discussed in Sections~\ref{Sec:SignalModel} and \ref{Sec:Background}. The common signal strength and the parameters of the background model are free parameters of the fit. The stability of the fit and the lack of bias with respect to the injected signals are validated with pseudo-experiments.

In addition to the statistical uncertainties arising from the size of each data set, systematic uncertainties are considered. The systematic uncertainties affecting the signal model can be divided into the rate uncertainties and the shape uncertainties. Both types of uncertainties are implemented in the signal model by means of nuisance parameters.

The JES and JER uncertainties affect the signal efficiency and the shape of the signal mass distribution, leading to a post-fit effect up to 5.8\%.
To estimate their influence, we evaluate the change of the distributions under variation of the energy and the resolution corrections for all jets within one standard deviation. Since the corrections are applied to the four-momentum of jets, they can affect the fraction of events passing the selection, which explains the normalization effect.

Ten separate uncertainties from the \PQb tagging simulation-to-data correction factors (scale factors) considering different jet flavors and run eras are studied. The shape variation effect caused by \PQb tagging is found to be negligible compared to that from JES and JER, hence only a rate uncertainty is applied. The impact of these uncertainties is less than 2\% for each uncertainty source. 

The estimate of the integrated luminosity leads to a cross section uncertainty in the range of 0.6--2\%~\cite{CMS:2021xjt,CMS-PAS-LUM-17-004,CMS-PAS-LUM-18-002}. There are five separate uncertainties, including three uncorrelated components corresponding to each of the data-taking years, one correlated component for all three years and one correlated component for 2017 and 2018. The overall uncertainty for the 2016--2018 data is 1.6\%. 

Lepton identification, isolation, and single-lepton trigger scale factors are rate uncertainties, where the scale factors are calculated for different working points and run periods and as a function of lepton \pt and $\eta$. The impact of each uncertainty is found to be less than 2\%. The trigger efficiency is estimated using a ``tag-and-probe" technique~\cite{CMS:2010vta}, based on the dileptons from the decay of a \PZ boson produced in a simulated DY background event.

The systematic uncertainty related to pileup is evaluated by varying the value of the total inelastic cross section by $\pm4.6\%$~\cite{CMS:2018mlc} in the pileup multiplicity calculation. The effect is up to 1\%.

During the 2016 and 2017 data-taking periods, a gradual shift in the timing of the inputs of the ECAL L1 trigger in the $\abs{\eta}>2.0$  region caused a trigger inefficiency, denoted as "prefiring"~\cite{CMS:2024ppo}. Correction factors were computed from data and applied to the acceptance evaluated by simulation, contributing as an additional rate uncertainty.

The PDF uncertainties are estimated using event-by-event weights corresponding to 100 replicas of the NNPDF set, following the PDF4LHC recommendations~\cite{Butterworth:2015oua}. In this analysis, only the contribution to the rate uncertainty is found to be relevant. The impact of this uncertainty is less than 3\%.  

Uncertainties in the renormalization and factorization scales ($\mu_{\text{R}}$ and $\mu_{\text{F}}$, respectively) used in the matrix-element calculations are accounted for as a rate effect, while their shape effect is found negligible. To quantify this effect, nine values in each event are varied independently by the factors of 0.5 and 2. Each value corresponds to the variation of $\mu_{\text{R}}$ or $\mu_{\text{F}}$, producing a separate weight. The standard deviation of the summed scale weights, excluding the values from the extreme variations, is taken as the uncertainty. The impact of this uncertainty is found to be less than 1\%.

Overall, the uncertainty is dominated by the effect of the nuisances related to the background parameterization and normalization, entirely constrained from the data, and therefore treated as statistical uncertainty. The effect of the systematic uncertainties on the signal on the exclusion limits is much smaller.

\section{Results}
\label{Sec:results}
As discussed in Section~\ref{Sec:Systematics}, a binned maximum likelihood fit of the signal plus background parameterizations to the \mtH distributions in the data is performed to extract the number of observed signal events. The results are ultimately combined by performing a joint fit of all categories using a single common signal strength parameter.

\begin{figure}
    \centering
    \includegraphics[width=0.45\textwidth]{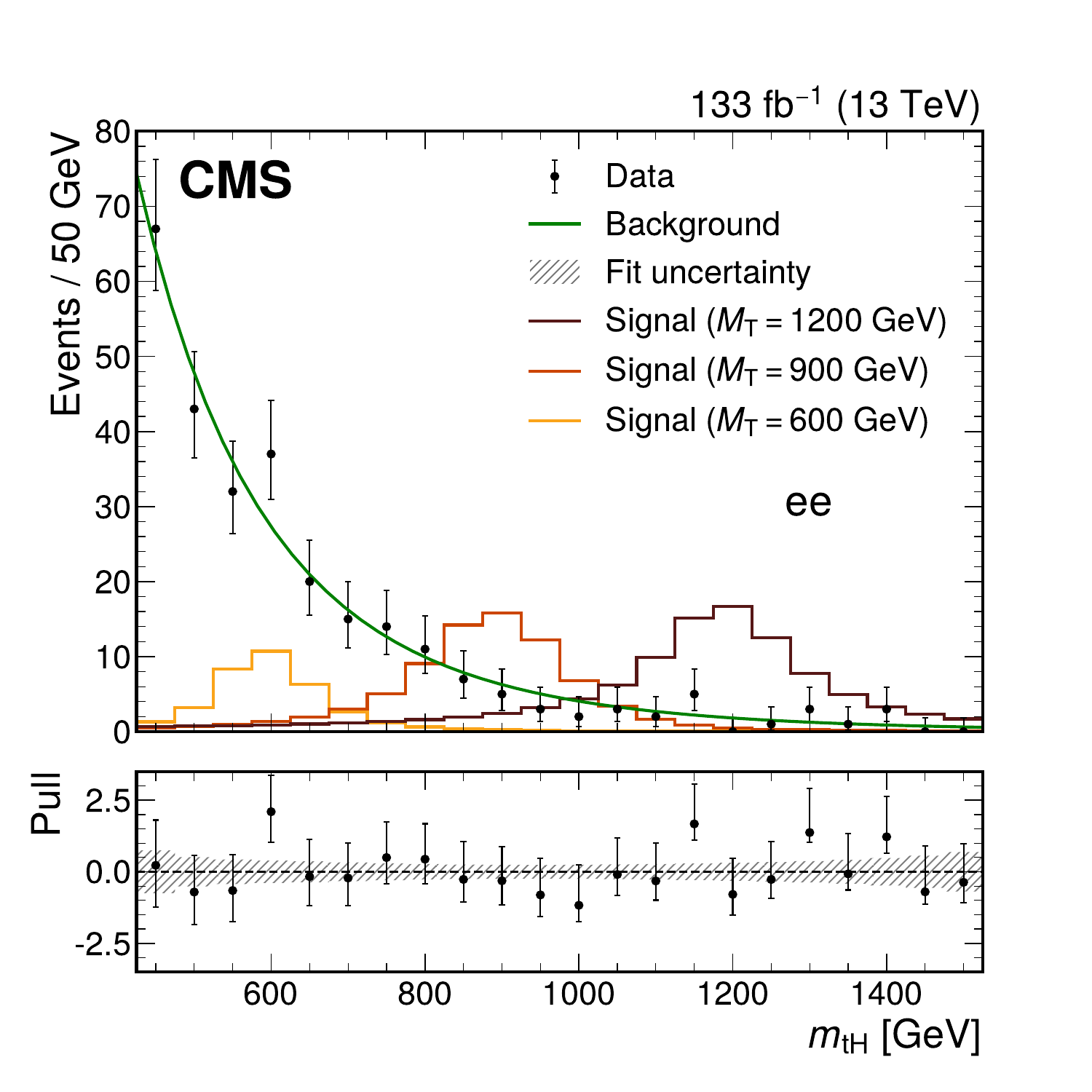}
    \includegraphics[width=0.45\textwidth]{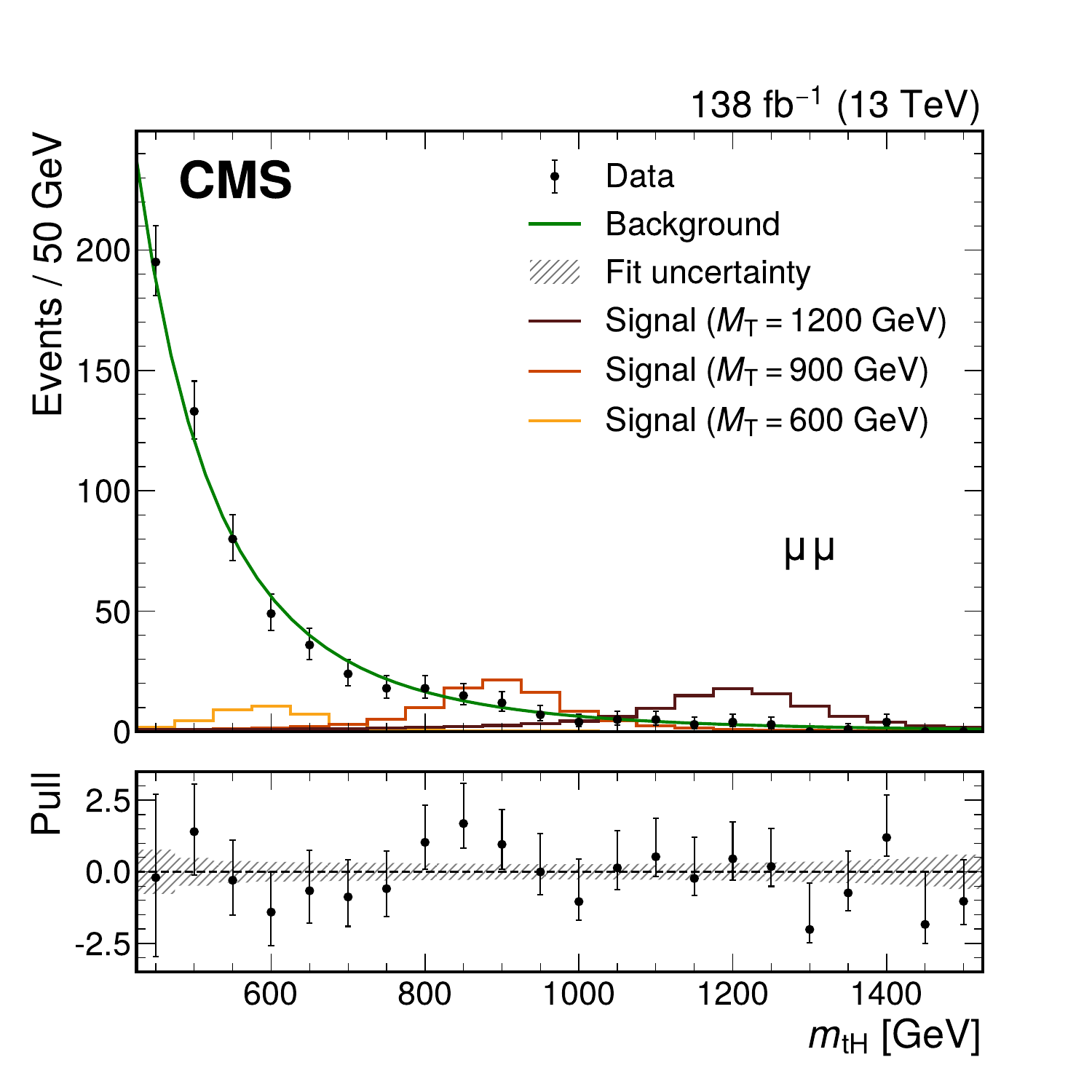}
    \includegraphics[width=0.45\textwidth]{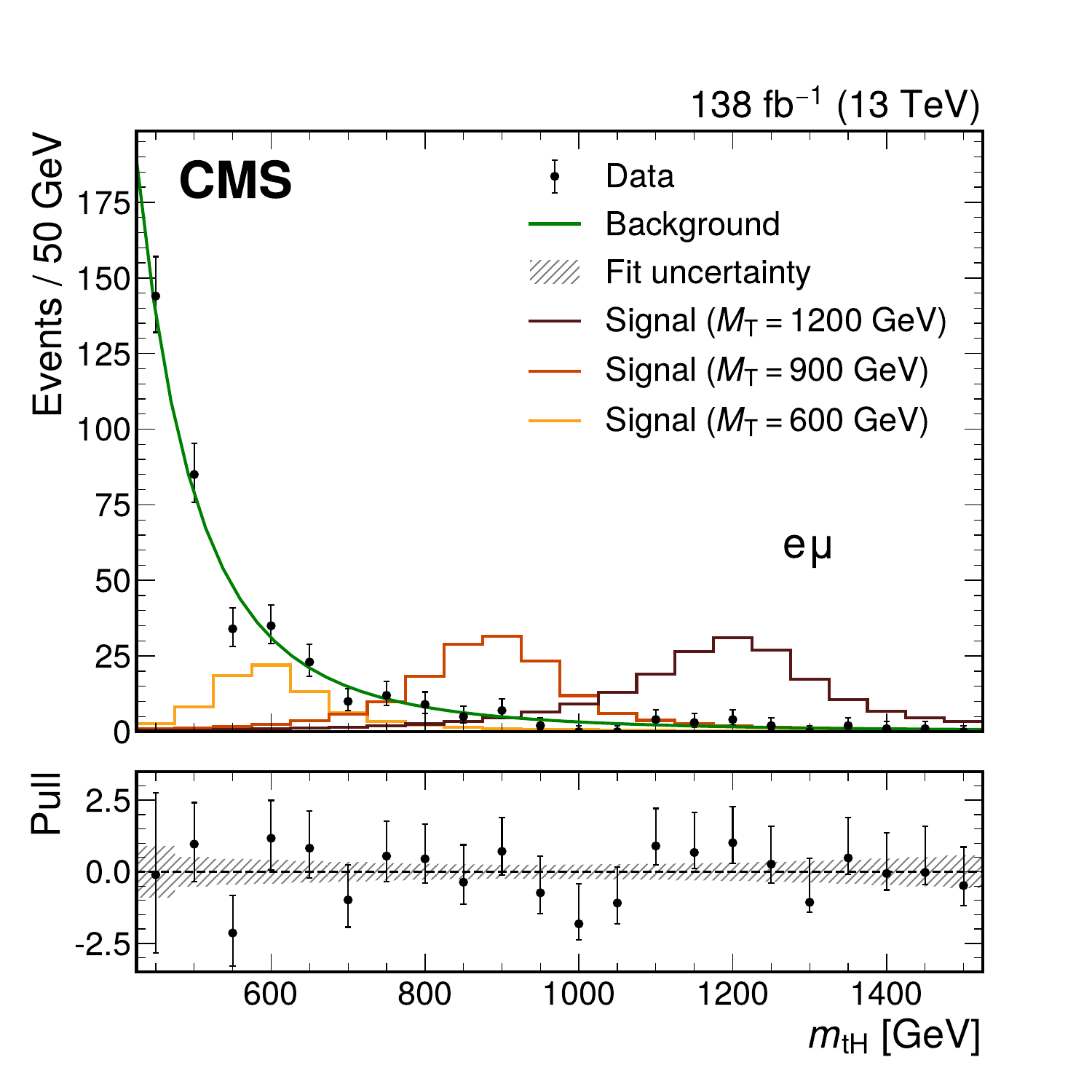}
    \hfill
    \caption{Distributions of the invariant mass \mtH in the  $\Pe\Pe$ (upper left),  $\Pgm\Pgm$ (upper right), and $\Pe\Pgm$ (lower) channels in 2016--2018 data set. In the upper panel, the black points show the data in the signal region, and the shapes from the background-only fits are in green. Expected signal distributions for three representative signal mass values are also shown, normalized to a cross section of 1\pb. The lower panel shows the pull, which is defined as $\frac{n_{\text{data}}-n_{\text{fit}}}{\sigma}$, where $\sigma^2 = \sigma^2_{\text{data}}-\sigma^2_{\text{fit}}$, and $\sigma_{\text{data}}$ (error bars) and $\sigma_{\text{fit}}$ (hatched bands) are statistical uncertainties of the data and of the fits, respectively. }
\label{fig:FitResult}
\end{figure}

The invariant mass distributions are shown in Fig.~\ref{fig:FitResult}, including the result of the background-only fits. In general, the data show very good agreement with the background-only hypothesis.

No significant excess of events is observed at any of the investigated mass points from 600 to 1200\GeV. This result is used to set 95\% confidence level (\CL) upper limits on the production cross section times branching fraction $\sigma(\Pp \Pp\to \PQT) {\mathcal{B}} (\PQT\to \PQt\PH)$, assuming the SM branching fractions for the underlying Higgs boson decay modes. The limits are computed using the modified frequentist \CLs criterion~\cite{Junk:1999kv, Read:2002hq}, making use of the asymptotic approximation~\cite{Cowan:2010js}. Systematic uncertainties are treated as nuisance parameters and profiled in the statistical inference using log-normal priors for uncertainties in the signal yield, and Gaussian priors for shape uncertainties.

\begin{figure}
    \centering
    \includegraphics[width=0.45\textwidth]{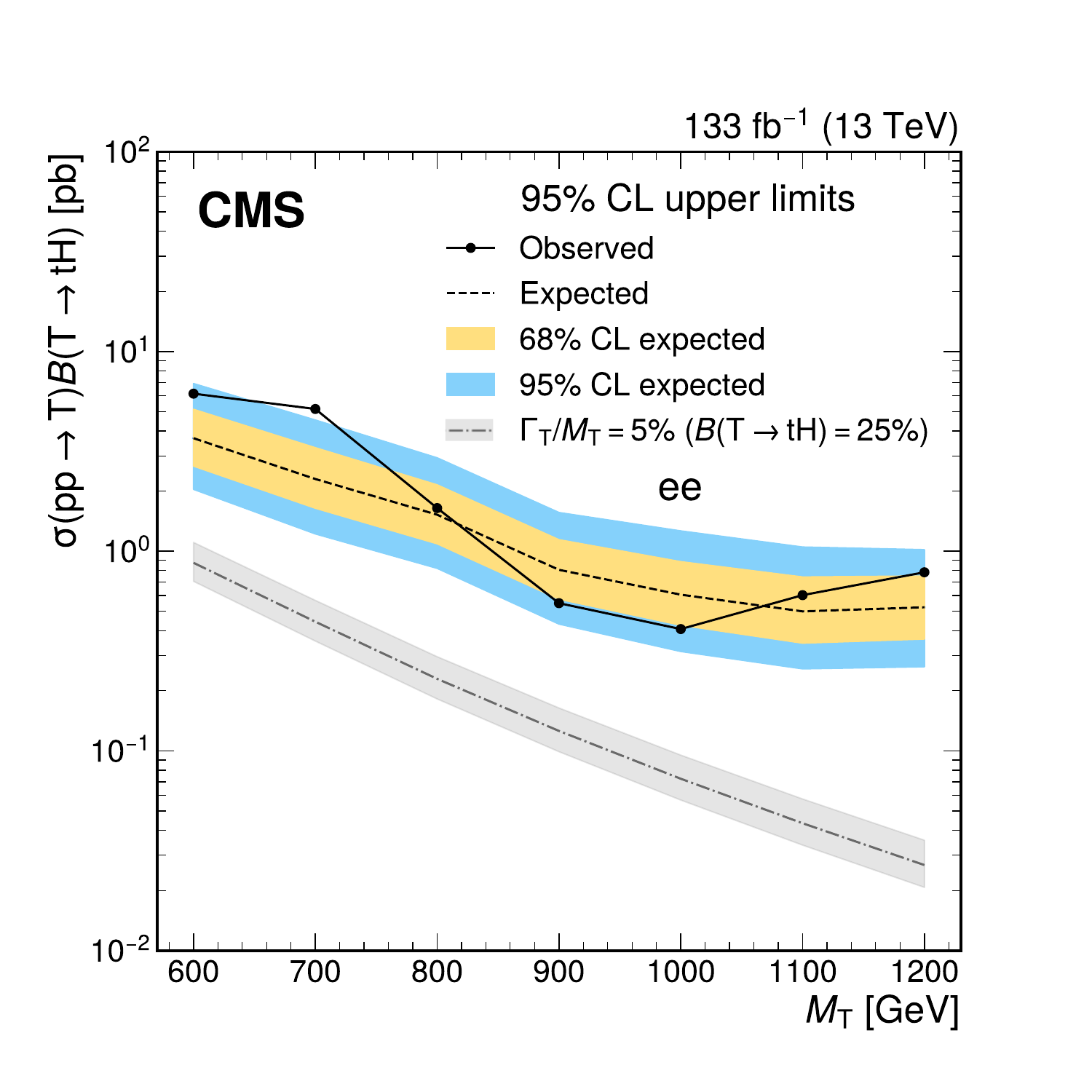}
    \includegraphics[width=0.45\textwidth]{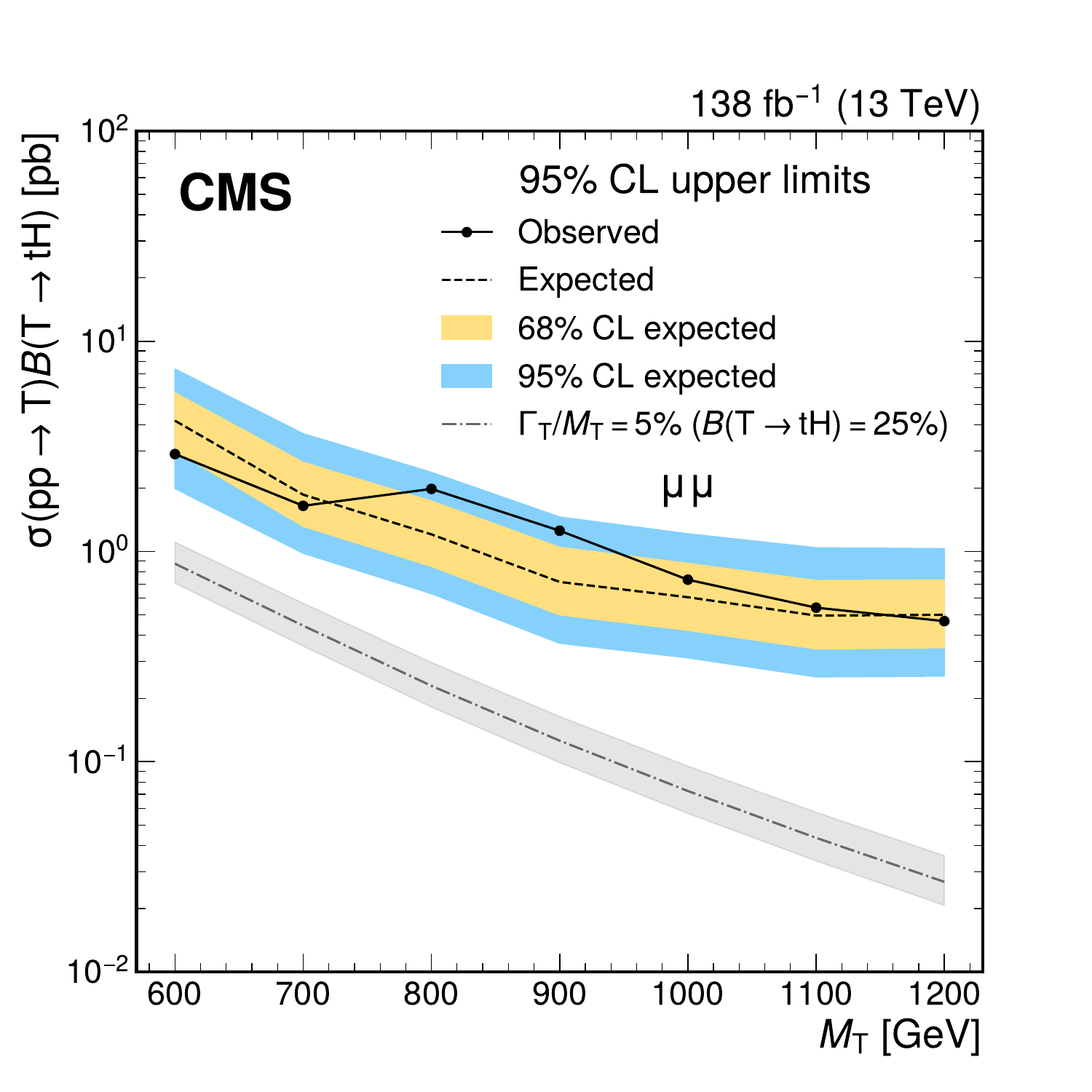}
    \includegraphics[width=0.45\textwidth]{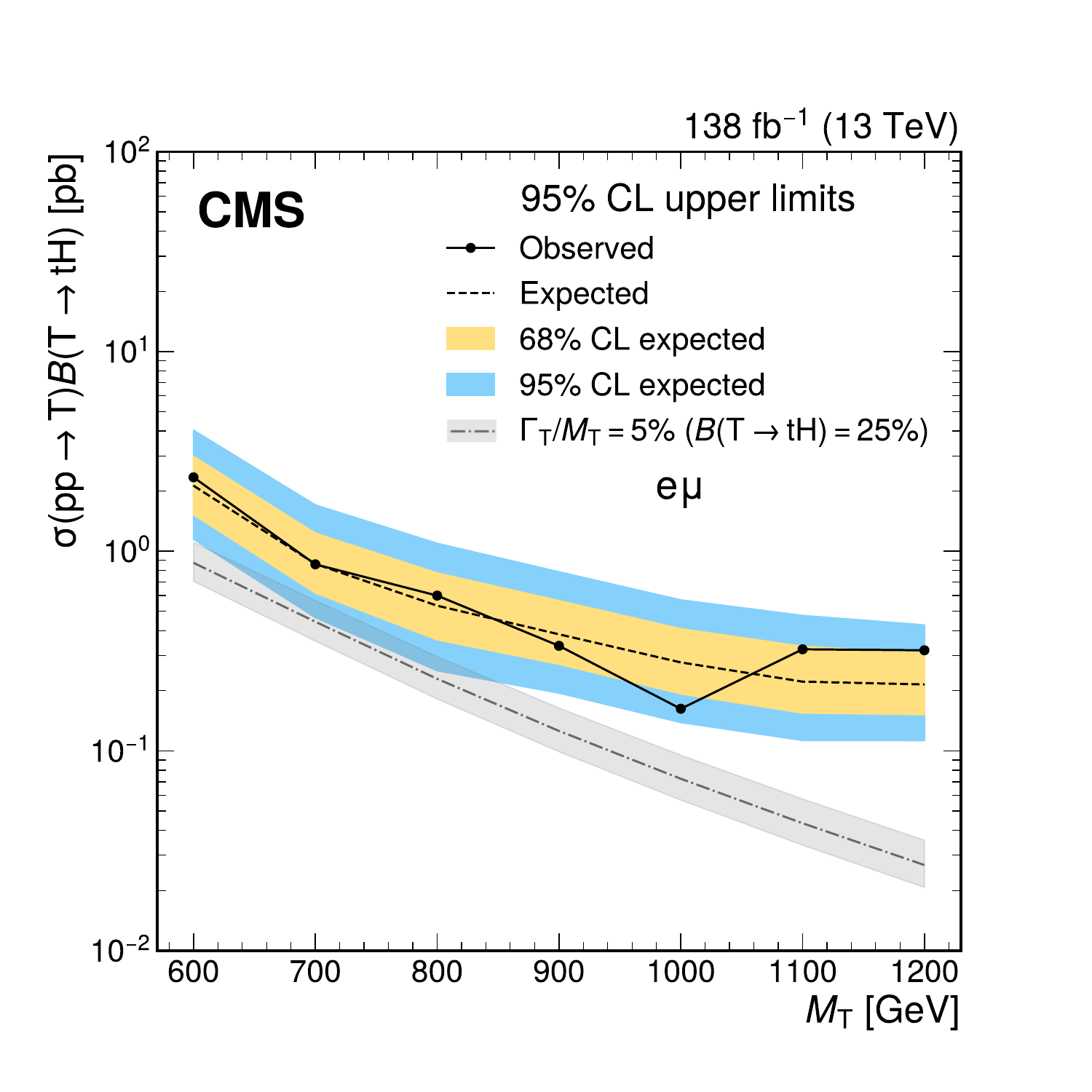}
    \caption{Expected and observed upper limits at 95\% \CL on the production cross section $\sigma(\Pp \Pp\to \PQT){\mathcal{B}}(\PQT \to \PQt\PH)$, for the three OS dilepton channels
$\Pe\Pe$ (upper left),  $\Pgm\Pgm$ (upper right), and $\Pe\Pgm$ (lower)
      in the 2016--2018 data set. The theoretical prediction assumes $\nwa=5\%$ and ${\mathcal{B}}(\PQT \to \PQt\PH) = 25\% $~\cite{Aguilar-Saavedra:2013qpa,Chen:2017hak}. The gray shaded band shows the uncertainty in the theory prediction.}
\label{fig:ExpectedLimit3channels}
\end{figure}

\begin{figure}
    \centering
    \includegraphics[width=0.8\textwidth]{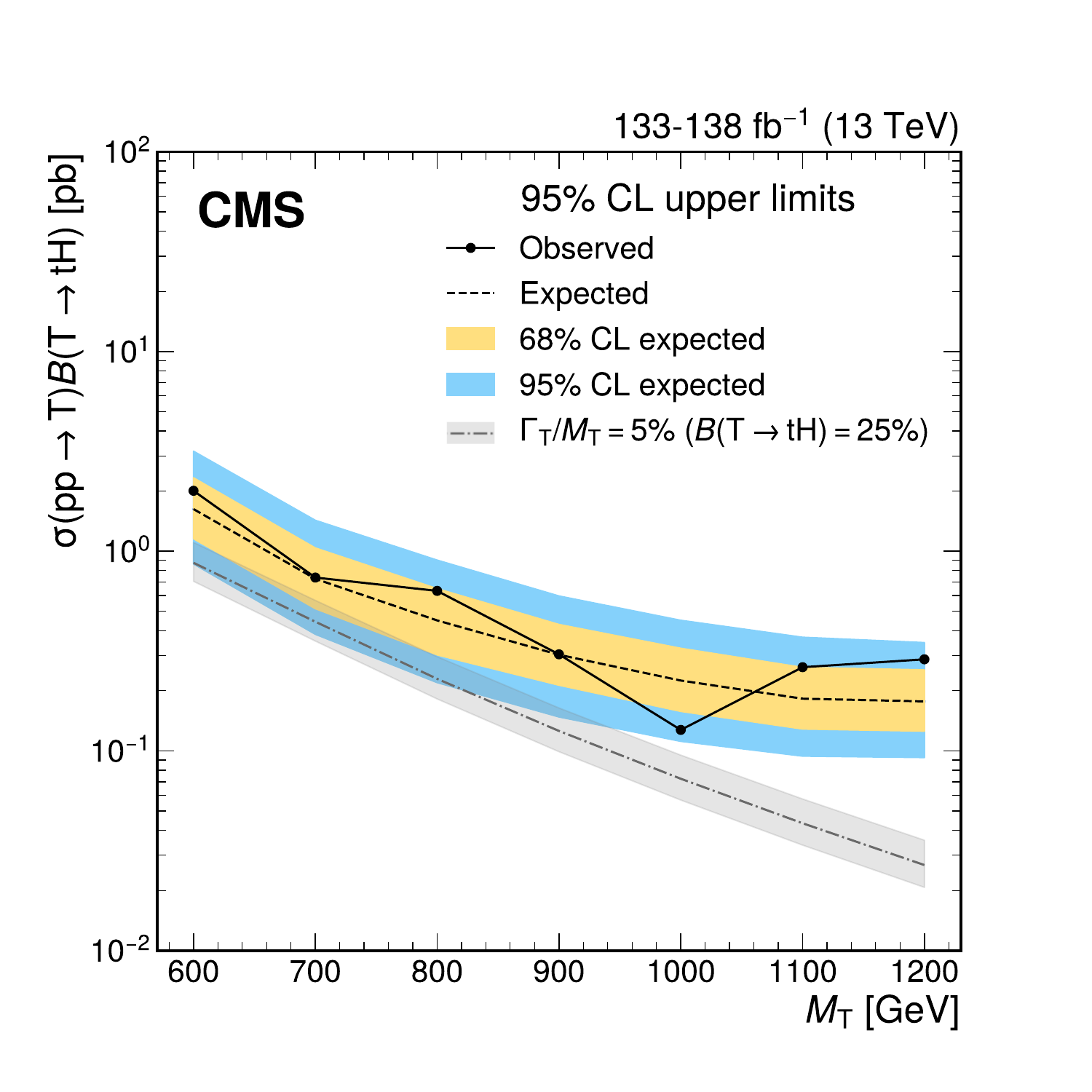}
    \hfill
    \caption{Expected and observed upper limits at 95\% \CL on the production cross section $\sigma(\Pp \Pp\to \PQT){\mathcal{B}}(\PQT \to \PQt\PH)$, combining all OS dilepton channels in the 2016--2018 data set. The theoretical prediction assumes $\nwa=5\%$ and ${\mathcal{B}}(\PQT \to \PQt\PH) = 25\% $~\cite{Aguilar-Saavedra:2013qpa,Chen:2017hak}. The gray shaded band shows the uncertainty in the theory prediction.}
\label{fig:ExpectedLimitRun2}
\end{figure}

Upper limits at 95\% \CL for the cross section $\sigma(\Pp \Pp\to \PQT) {\mathcal{B}}(\PQT \to \PQt\PH)$ are shown in Fig.~\ref{fig:ExpectedLimit3channels} for each channel, and in Fig.~\ref{fig:ExpectedLimitRun2} for the combination of all three dilepton channels in the 2016--2018 data. A theory prediction of the LO cross section of the process $\Pp \Pp \to \PQT \PQb \PQq $ assuming $\nwa=5\%$ and ${\mathcal{B}} (\PQT\to \PQt\PH)=25\%$ is superimposed. The combined limits range from 2.0\pb at a mass of 600\GeV to about 0.1\pb at 1000\GeV. The observed limits are in good agreement with the expected limits within the uncertainty bands. The strongest limits are obtained from the $\Pe\Pgm$ channel because it has the highest branching fraction and an absence of DY background.

This search exploits the signature with OS dileptons for the first time and complements previous searches for single \PQT VLQ production in Ref.~\cite{CMS:2024bni}.

\clearpage

\section{Summary}
\label{sec:Summary}
A search for singly produced vector-like \PQT quark decaying into the standard model top quark and Higgs boson has been presented, performed with the 2016--2018 data collected by the CMS experiment in proton-proton collisions at a center-of-mass energy of 13\TeV, corresponding to an integrated luminosity of up to 138\fbinv. The mass range from 600 to 1200\GeV is probed. No significant excess above the expected standard model background is observed.
The results are used to set upper limits at 95\% confidence level on the product of the production cross section and the decay branching fraction. The observed limits are 2.0\pb at a \PQT mass $\MT=600\GeV$ and 0.3\pb at $\MT=1200\GeV$, with a minimum of 0.1\pb at $\MT=1000\GeV$. Single vector-like \PQT quark production, followed by a decay to a top quark and a Higgs boson, in final states with two oppositely charged electrons or muons, is probed for the first time by this search.

\begin{acknowledgments}
We congratulate our colleagues in the CERN accelerator departments for the excellent performance of the LHC and thank the technical and administrative staffs at CERN and at other CMS institutes for their contributions to the success of the CMS effort. In addition, we gratefully acknowledge the computing centers and personnel of the Worldwide LHC Computing Grid and other centers for delivering so effectively the computing infrastructure essential to our analyses. Finally, we acknowledge the enduring support for the construction and operation of the LHC, the CMS detector, and the supporting computing infrastructure provided by the following funding agencies: SC (Armenia), BMBWF and FWF (Austria); FNRS and FWO (Belgium); CNPq, CAPES, FAPERJ, FAPERGS, and FAPESP (Brazil); MES and BNSF (Bulgaria); CERN; CAS, MoST, and NSFC (China); MINCIENCIAS (Colombia); MSES and CSF (Croatia); RIF (Cyprus); SENESCYT (Ecuador); ERC PRG and PSG, TARISTU24-TK10 and MoER TK202 (Estonia); Academy of Finland, MEC, and HIP (Finland); CEA and CNRS/IN2P3 (France); SRNSF (Georgia); BMFTR, DFG, and HGF (Germany); GSRI (Greece); MATE and NKFIH (Hungary); DAE and DST (India); IPM (Iran); SFI (Ireland); INFN (Italy); MSIT and NRF (Republic of Korea); MES (Latvia); LMTLT (Lithuania); MOE and UM (Malaysia); BUAP, CINVESTAV, CONACYT, LNS, SEP, and UASLP-FAI (Mexico); MOS (Montenegro); MBIE (New Zealand); PAEC (Pakistan); MSHE, NSC, and NAWA (Poland); FCT (Portugal); MESTD (Serbia); MICIU/AEI and PCTI (Spain); MOSTR (Sri Lanka); Swiss Funding Agencies (Switzerland); MST (Taipei); MHESI (Thailand); TUBITAK and TENMAK (T\"{u}rkiye); NASU (Ukraine); STFC (United Kingdom); DOE and NSF (USA).

\begin{sloppypar}
\setlength\emergencystretch{\hsize}
\hyphenation{Rachada-pisek} Individuals have received support from the Marie-Curie programme and the European Research Council and Horizon 2020 Grant, contract Nos.\ 675440, 724704, 752730, 758316, 765710, 824093, 101115353, 101002207, 101001205, and COST Action CA16108 (European Union); the Leventis Foundation; the Alfred P.\ Sloan Foundation; the Alexander von Humboldt Foundation; the Science Committee, project no. 22rl-037 (Armenia); the Fonds pour la Formation \`a la Recherche dans l'Industrie et dans l'Agriculture (FRIA) and Fonds voor Wetenschappelijk Onderzoek contract No. 1228724N (Belgium); the Beijing Municipal Science \& Technology Commission, No. Z191100007219010, the Fundamental Research Funds for the Central Universities, the Ministry of Science and Technology of China under Grant No. 2023YFA1605804, the Natural Science Foundation of China under Grant No. 12535004, and USTC Research Funds of the Double First-Class Initiative No.\ YD2030002017 (China); the Ministry of Education, Youth and Sports (MEYS) of the Czech Republic; the Shota Rustaveli National Science Foundation (Georgia); the Deutsche Forschungsgemeinschaft (DFG), among others, under Germany's Excellence Strategy -- EXC 2121 ``Quantum Universe" -- 390833306, and under project number 400140256 - GRK2497; the Hellenic Foundation for Research and Innovation (HFRI), Project Number 2288 (Greece); the Hungarian Academy of Sciences, the New National Excellence Program - \'UNKP, the NKFIH research grants K 131991, K 138136, K 143460, K 143477, K 147557, K 146913, K 146914, K 147048, TKP2021-NKTA-64, and 2025-1.1.5-NEMZ\_KI-2025-00004, and MATE KKP and KKPCs Research Excellence and Flagship Research Groups grants (Hungary); the Council of Science and Industrial Research, India; ICSC -- National Research Center for High Performance Computing, Big Data and Quantum Computing, FAIR -- Future Artificial Intelligence Research, and CUP I53D23001070006 (Mission 4 Component 1), funded by the NextGenerationEU program, the Italian Ministry of University and Research (MUR) under Bando PRIN 2022 -- CUP I53C24002390006, PRIN PRIMULA 2022RBYK7T (Italy); the Latvian Council of Science; the Ministry of Science and Higher Education, project no. 2022/WK/14, and the National Science Center, contracts Opus 2021/41/B/ST2/01369, 2021/43/B/ST2/01552, 2023/49/B/ST2/03273, and the NAWA contract BPN/PPO/2021/1/00011 (Poland); the Funda\c{c}\~ao para a Ci\^encia e a Tecnologia (Portugal); the National Priorities Research Program by Qatar National Research Fund; MICIU/AEI/10.13039/501100011033, ERDF/EU, ``European Union NextGenerationEU/PRTR", projects PID2022-142604OB-C21, PID2022-139519OB-C21, PID2023-147706NB-I00, PID2023-148896NB-I00, PID2023-146983NB-I00, PID2023-147115NB-I00, PID2023-148418NB-C41, PID2023-148418NB-C42, PID2023-148418NB-C43, PID2023-148418NB-C44, PID2024-158190NB-C22, RYC2021-033305-I, RYC2024-048719-I, CNS2023-144781, CNS2024-154769 and Plan de Ciencia, Tecnolog{\'i}a e Innovaci{\'o}n de Asturias, Spain; the Chulalongkorn Academic into Its 2nd Century Project Advancement Project, the National Science, Research and Innovation Fund program IND\_FF\_68\_369\_2300\_097, and the Program Management Unit for Human Resources \& Institutional Development, Research and Innovation, grant B39G680009 (Thailand); the Eric \& Wendy Schmidt Fund for Strategic Innovation through the CERN Next Generation Triggers project under grant agreement number SIF-2023-004; the Kavli Foundation; the Nvidia Corporation; the SuperMicro Corporation; the Welch Foundation, contract C-1845; and the Weston Havens Foundation (USA).
\end{sloppypar}
\end{acknowledgments}\section*{Data availability} Release and preservation of data used by the CMS Collaboration as the basis for publications is guided by the  \href{https://doi.org/10.7483/OPENDATA.CMS.1BNU.8V1W}{CMS data preservation, re-use and open access policy}.

\bibliography{auto_generated}

\providecommand{\href}[2]{#2}\begingroup\raggedright\begin{thebibliography}{10}%
\makeatletter
\providecommand{\hrefCMSnoop }[0]{\@secondoftwo}%
\makeatother
\providecommand{\doi}{\texttt{doi:}\begingroup \urlstyle{tt}\Url}

\bibitem{Theory}
\hrefCMSnoop {}{A.~Djouadi and A.~Lenz, ``Sealing the fate of a fourth
  generation of fermions'',} \textit{ Phys. Lett. B} \textbf{ 715} (2012) 310,
  \href{http://dx.doi.org/10.1016/j.physletb.2012.07.060}{\doi{10.1016/j.physletb.2012.07.060}},
  \href{http://www.arXiv.org/abs/1204.1252}{\texttt{arXiv:1204.1252}}.

\bibitem{Lenz:2013iha}
\hrefCMSnoop {}{A.~Lenz, ``Constraints on a fourth generation of fermions from
  {H}iggs boson searches'',} \textit{ Adv. High Energy Phys.} \textbf{ 2013}
  (2013) 910275,
  \href{http://dx.doi.org/10.1155/2013/910275}{\doi{10.1155/2013/910275}}.

\bibitem{Branco:2021vhs}
G.~C. Branco\hrefCMSnoop {}{ { et~al.}, ``Addressing the {CKM} unitarity
  problem with a vector-like up quark'',} \textit{ JHEP} \textbf{ 07} (2021)
  099,
  \href{http://dx.doi.org/10.1007/JHEP07(2021)099}{\doi{10.1007/JHEP07(2021)099}},
  \href{http://www.arXiv.org/abs/2103.13409}{\texttt{arXiv:2103.13409}}.

\bibitem{Belfatto:2023tbv}
\hrefCMSnoop {}{B.~Belfatto and S.~Trifinopoulos, ``Cabibbo angle anomalies and
  oblique corrections: {T}he remarkable role of the vectorlike quark
  doublet'',} \textit{ Phys. Rev. D} \textbf{ 108} (2023) 035022,
  \href{http://dx.doi.org/10.1103/PhysRevD.108.035022}{\doi{10.1103/PhysRevD.108.035022}},
  \href{http://www.arXiv.org/abs/2302.14097}{\texttt{arXiv:2302.14097}}.

\bibitem{Pocanic:2003pf}
\hrefCMSnoop {}{D.~Pocanic { et~al.}, ``Precise measurement of the {$\Pgpp \to
  \Pgpz \Pep \Pgn$} branching ratio'',} \textit{ Phys. Rev. Lett.} \textbf{ 93}
  (2004) 181803,
  \href{http://dx.doi.org/10.1103/PhysRevLett.93.181803}{\doi{10.1103/PhysRevLett.93.181803}},
  \href{http://www.arXiv.org/abs/hep-ex/0312030}{\texttt{arXiv:hep-ex/0312030}}.

\bibitem{Czarnecki:2019iwz}
\hrefCMSnoop {}{A.~Czarnecki, W.~J. Marciano, and A.~Sirlin, ``Pion beta decay
  and {Cabibbo-Kobayashi-Maskawa} unitarity'',} \textit{ Phys. Rev. D} \textbf{
  101} (2020) 091301,
  \href{http://dx.doi.org/10.1103/PhysRevD.101.091301}{\doi{10.1103/PhysRevD.101.091301}},
  \href{http://www.arXiv.org/abs/1911.04685}{\texttt{arXiv:1911.04685}}.

\bibitem{Aguilar-Saavedra:2013qpa}
\hrefCMSnoop {}{J.~A. Aguilar-Saavedra, R.~Benbrik, S.~Heinemeyer, and
  M.~P{\'e}rez-Victoria, ``Handbook of vectorlike quarks: {M}ixing and single
  production'',} \textit{ Phys. Rev. D} \textbf{ 88} (2013) 094010,
  \href{http://dx.doi.org/10.1103/PhysRevD.88.094010}{\doi{10.1103/PhysRevD.88.094010}},
  \href{http://www.arXiv.org/abs/1306.0572}{\texttt{arXiv:1306.0572}}.

\bibitem{ATLAS:2024xdc}
\hrefCMSnoop {}{{ATLAS Collaboration}, ``Combination of searches for singly
  produced vectorlike top quarks in pp collisions at {$ \sqrt{s}=13 $ TeV} with
  the {ATLAS} detector'',} \textit{ Phys. Rev. D} \textbf{ 111} (2025) 012012,
  \href{http://dx.doi.org/10.1103/PhysRevD.111.012012}{\doi{10.1103/PhysRevD.111.012012}},
  \href{http://www.arXiv.org/abs/2408.08789}{\texttt{arXiv:2408.08789}}.

\bibitem{CMS:2024bni}
\hrefCMSnoop {}{{CMS Collaboration}, ``Review of searches for vector-like
  quarks, vector-like leptons, and heavy neutral leptons in
  proton{\textendash}proton collisions at {$\sqrt{s} = 13~\mathrm{TeV}$} at the
  {CMS} experiment'',} \textit{ Phys. Rept.} \textbf{ 1115} (2025) 570,
  \href{http://dx.doi.org/10.1016/j.physrep.2024.09.012}{\doi{10.1016/j.physrep.2024.09.012}},
  \href{http://www.arXiv.org/abs/2405.17605}{\texttt{arXiv:2405.17605}}.

\bibitem{Aguilar-Saavedra:2009xmz}
\hrefCMSnoop {}{J.~A. Aguilar-Saavedra, ``Identifying top partners at {LHC}'',}
  \textit{ JHEP} \textbf{ 11} (2009) 030,
  \href{http://dx.doi.org/10.1088/1126-6708/2009/11/030}{\doi{10.1088/1126-6708/2009/11/030}},
  \href{http://www.arXiv.org/abs/0907.3155}{\texttt{arXiv:0907.3155}}.

\bibitem{CMS:2019afi}
\hrefCMSnoop {}{{CMS Collaboration}, ``Search for electroweak production of a
  vector-like {T} quark using fully hadronic final states'',} \textit{ JHEP}
  \textbf{ 01} (2020) 036,
  \href{http://dx.doi.org/10.1007/JHEP01(2020)036}{\doi{10.1007/JHEP01(2020)036}},
  \href{http://www.arXiv.org/abs/1909.04721}{\texttt{arXiv:1909.04721}}.

\bibitem{CMS:2023agg}
\hrefCMSnoop {}{{CMS Collaboration}, ``Search for a vector-like quark
  {T$'$$\to$ tH} via the diphoton decay mode of the {H}iggs boson in
  proton-proton collisions at {$\sqrt{s}$ = 13 TeV}'',} \textit{ JHEP} \textbf{
  09} (2023) 057,
  \href{http://dx.doi.org/10.1007/JHEP09(2023)057}{\doi{10.1007/JHEP09(2023)057}},
  \href{http://www.arXiv.org/abs/2302.12802}{\texttt{arXiv:2302.12802}}.

\bibitem{CMS:2024qdd}
\hrefCMSnoop {}{{CMS Collaboration}, ``Search for production of a single
  vector-like quark decaying to t{H} or t{Z} in the all-hadronic final state in
  pp collisions at {$\sqrt{s} = 13~\mathrm{TeV}$}'',} \textit{ Phys. Rev. D}
  \textbf{ 110} (2024) 072012,
  \href{http://dx.doi.org/10.1103/PhysRevD.110.072012}{\doi{10.1103/PhysRevD.110.072012}},
  \href{http://www.arXiv.org/abs/2405.05071}{\texttt{arXiv:2405.05071}}.

\bibitem{HEPData}
\hrefCMSnoop {}{}{HEPD}ata record for this analysis, 2026.
\newblock
  \href{http://dx.doi.org/10.17182/hepdata.166760}{\doi{10.17182/hepdata.166760}}.

\bibitem{CMS:2008xjf}
\hrefCMSnoop {}{{CMS Collaboration}, ``The {CMS} experiment at the {CERN}
  {LHC}'',} \textit{ JINST} \textbf{ 3} (2008) S08004,
  \href{http://dx.doi.org/10.1088/1748-0221/3/08/S08004}{\doi{10.1088/1748-0221/3/08/S08004}}.

\bibitem{CMS:2023gfb}
\hrefCMSnoop {}{{CMS Collaboration}, ``Development of the {CMS} detector for
  the {CERN} {LHC} {R}un 3'',} \textit{ JINST} \textbf{ 19} (2024) P05064,
  \href{http://dx.doi.org/10.1088/1748-0221/19/05/P05064}{\doi{10.1088/1748-0221/19/05/P05064}},
  \href{http://www.arXiv.org/abs/2309.05466}{\texttt{arXiv:2309.05466}}.

\bibitem{CMS:2016ngn}
\hrefCMSnoop {}{{CMS Collaboration}, ``The {CMS} trigger system'',} \textit{
  JINST} \textbf{ 12} (2017) P01020,
  \href{http://dx.doi.org/10.1088/1748-0221/12/01/P01020}{\doi{10.1088/1748-0221/12/01/P01020}},
\href{http://www.arXiv.org/abs/1609.02366}{\texttt{arXiv:1609.02366}}.
%%CITATION = ARXIV:1609.02366;%%.

\bibitem{CMS:2020cmk}
\hrefCMSnoop {}{{CMS Collaboration}, ``Performance of the {CMS} {L}evel-1
  trigger in proton-proton collisions at {$\sqrt{s} = 13$\,TeV}'',} \textit{
  JINST} \textbf{ 15} (2020) P10017,
  \href{http://dx.doi.org/10.1088/1748-0221/15/10/P10017}{\doi{10.1088/1748-0221/15/10/P10017}},
  \href{http://www.arXiv.org/abs/2006.10165}{\texttt{arXiv:2006.10165}}.

\bibitem{CMS:2024aqx}
\hrefCMSnoop {}{{CMS Collaboration}, ``Performance of the {CMS} high-level
  trigger during {LHC} {Run} 2'',} \textit{ JINST} \textbf{ 19} (2024) P11021,
  \href{http://dx.doi.org/10.1088/1748-0221/19/11/P11021}{\doi{10.1088/1748-0221/19/11/P11021}},
  \href{http://www.arXiv.org/abs/2410.17038}{\texttt{arXiv:2410.17038}}.

\bibitem{CMS:2017yfk}
\hrefCMSnoop {}{{CMS Collaboration}, ``Particle-flow reconstruction and global
  event description with the {CMS} detector'',} \textit{ JINST} \textbf{ 12}
  (2017) P10003,
  \href{http://dx.doi.org/10.1088/1748-0221/12/10/P10003}{\doi{10.1088/1748-0221/12/10/P10003}},
\href{http://www.arXiv.org/abs/1706.04965}{\texttt{arXiv:1706.04965}}.
%%CITATION = ARXIV:1706.04965;%%.

\bibitem{CMS:2014pgm}
\hrefCMSnoop {}{{CMS Collaboration}, ``Description and performance of track and
  primary-vertex reconstruction with the {CMS} tracker'',} \textit{ JINST}
  \textbf{ 9} (2014) P10009,
  \href{http://dx.doi.org/10.1088/1748-0221/9/10/P10009}{\doi{10.1088/1748-0221/9/10/P10009}},
  \href{http://www.arXiv.org/abs/1405.6569}{\texttt{arXiv:1405.6569}}.

\bibitem{Cacciari:2008gp}
\hrefCMSnoop {}{M.~Cacciari, G.~P. Salam, and G.~Soyez, ``{The anti-\kt} jet
  clustering algorithm'',} \textit{ JHEP} \textbf{ 04} (2008) 063,
  \href{http://dx.doi.org/10.1088/1126-6708/2008/04/063}{\doi{10.1088/1126-6708/2008/04/063}},
  \href{http://www.arXiv.org/abs/0802.1189}{\texttt{arXiv:0802.1189}}.

\bibitem{Cacciari:2011ma}
\hrefCMSnoop {}{M.~Cacciari, G.~P. Salam, and G.~Soyez, ``Fastjet user
  manual'',} \textit{ Eur. Phys. J. C} \textbf{ 72} (2012) 1896,
  \href{http://dx.doi.org/10.1140/epjc/s10052-012-1896-2}{\doi{10.1140/epjc/s10052-012-1896-2}},
\href{http://www.arXiv.org/abs/1111.6097}{\texttt{arXiv:1111.6097}}.
%%CITATION = ARXIV:1111.6097;%%.

\bibitem{CMS:2016lmd}
\hrefCMSnoop {}{{CMS Collaboration}, ``Jet energy scale and resolution in the
  {CMS} experiment in pp collisions at 8 {TeV}'',} \textit{ JINST} \textbf{ 12}
  (2017) P02014,
  \href{http://dx.doi.org/10.1088/1748-0221/12/02/P02014}{\doi{10.1088/1748-0221/12/02/P02014}},
\href{http://www.arXiv.org/abs/1607.03663}{\texttt{arXiv:1607.03663}}.
%%CITATION = ARXIV:1607.03663;%%.

\bibitem{CMS:2020ebo}
\hrefCMSnoop {}{{CMS Collaboration}, ``Pileup mitigation at {CMS} in 13 {TeV}
  data'',} \textit{ JINST} \textbf{ 15} (2020) P09018,
  \href{http://dx.doi.org/10.1088/1748-0221/15/09/P09018}{\doi{10.1088/1748-0221/15/09/P09018}},
  \href{http://www.arXiv.org/abs/2003.00503}{\texttt{arXiv:2003.00503}}.

\bibitem{CMS:2019ctu}
\hrefCMSnoop {}{{CMS Collaboration}, ``Performance of missing transverse
  momentum reconstruction in proton-proton collisions at {$\sqrt{s} =
  13$}\,{TeV} using the {CMS} detector'',} \textit{ JINST} \textbf{ 14} (2019)
  P07004,
  \href{http://dx.doi.org/10.1088/1748-0221/14/07/P07004}{\doi{10.1088/1748-0221/14/07/P07004}},
\href{http://www.arXiv.org/abs/1903.06078}{\texttt{arXiv:1903.06078}}.
%%CITATION = ARXIV:1903.06078;%%.

\bibitem{Bols:2020bkb}
E.~Bols\hrefCMSnoop {}{ { et~al.}, ``Jet flavour classification using
  {\textsc{deepjet}}'',} \textit{ JINST} \textbf{ 15} (2020) P12012,
  \href{http://dx.doi.org/10.1088/1748-0221/15/12/P12012}{\doi{10.1088/1748-0221/15/12/P12012}},
  \href{http://www.arXiv.org/abs/2008.10519}{\texttt{arXiv:2008.10519}}.

\bibitem{CMS-DP-2023-005}
\href {https://cds.cern.ch/record/2854609}{{CMS Collaboration}, ``Performance
  summary of {AK4} jet b tagging with data from proton-proton collisions at 13
  {TeV} with the {CMS} detector'',} CMS Detector Performance Summary
  CMS-DP-2023-005, 2023.

\bibitem{CMS:2017wtu}
\hrefCMSnoop {}{{CMS Collaboration}, ``Identification of heavy-flavour jets
  with the {CMS} detector in pp collisions at 13 {TeV}'',} \textit{ JINST}
  \textbf{ 13} (2018) P05011,
  \href{http://dx.doi.org/10.1088/1748-0221/13/05/P05011}{\doi{10.1088/1748-0221/13/05/P05011}},
  \href{http://www.arXiv.org/abs/1712.07158}{\texttt{arXiv:1712.07158}}.

\bibitem{CMS-DP-2018-058}
\href {http://cds.cern.ch/record/2646773}{{CMS Collaboration}, ``Performance of
  the {\textsc{deepjet}} b tagging algorithm using 41.9 {$\mathrm{fb^{-1}}$} of
  data from proton-proton collisions at 13 {TeV} with {P}hase 1 {CMS}
  detector'',} CMS Detector Performance Summary CMS-DP-2018-058, 2018.

\bibitem{Alwall:2014hca}
J.~Alwall\hrefCMSnoop {}{ { et~al.}, ``The automated computation of tree-level
  and next-to-leading order differential cross sections, and their matching to
  parton shower simulations'',} \textit{ JHEP} \textbf{ 07} (2014) 079,
  \href{http://dx.doi.org/10.1007/JHEP07(2014)079}{\doi{10.1007/JHEP07(2014)079}},
  \href{http://www.arXiv.org/abs/1405.0301}{\texttt{arXiv:1405.0301}}.

\bibitem{Alioli:2010xd}
\hrefCMSnoop {}{S.~Alioli, P.~Nason, C.~Oleari, and E.~Re, ``A general
  framework for implementing {NLO} calculations in shower {M}onte {C}arlo
  programs: the {POWHEG BOX}'',} \textit{ JHEP} \textbf{ 06} (2010) 043,
  \href{http://dx.doi.org/10.1007/JHEP06(2010)043}{\doi{10.1007/JHEP06(2010)043}},
  \href{http://www.arXiv.org/abs/1002.2581}{\texttt{arXiv:1002.2581}}.

\bibitem{Frixione:2002ik}
\hrefCMSnoop {}{S.~Frixione and B.~R. Webber, ``Matching {NLO} {QCD}
  computations and parton shower simulations'',} \textit{ JHEP} \textbf{ 06}
  (2002) 029,
  \href{http://dx.doi.org/10.1088/1126-6708/2002/06/029}{\doi{10.1088/1126-6708/2002/06/029}},
  \href{http://www.arXiv.org/abs/hep-ph/0204244}{\texttt{arXiv:hep-ph/0204244}}.

\bibitem{NNPDF:2017mvq}
\hrefCMSnoop {}{{NNPDF} Collaboration, ``Parton distributions from
  high-precision collider data'',} \textit{ Eur. Phys. J. C} \textbf{ 77}
  (2017) 663,
  \href{http://dx.doi.org/10.1140/epjc/s10052-017-5199-5}{\doi{10.1140/epjc/s10052-017-5199-5}},
  \href{http://www.arXiv.org/abs/1706.00428}{\texttt{arXiv:1706.00428}}.

\bibitem{Sjostrand:2014zea}
T.~Sj{\"o}strand\hrefCMSnoop {}{ { et~al.}, ``An introduction to {\PYTHIA}
  8.2'',} \textit{ Comput. Phys. Commun.} \textbf{ 191} (2015) 159,
  \href{http://dx.doi.org/10.1016/j.cpc.2015.01.024}{\doi{10.1016/j.cpc.2015.01.024}},
  \href{http://www.arXiv.org/abs/1410.3012}{\texttt{arXiv:1410.3012}}.

\bibitem{GEANT4:2002zbu}
\hrefCMSnoop {}{{\GEANTfour} Collaboration, ``{\GEANTfour ---} a simulation
  toolkit'',} \textit{ Nucl. Instrum. Meth. A} \textbf{ 506} (2003) 250,
  \href{http://dx.doi.org/10.1016/S0168-9002(03)01368-8}{\doi{10.1016/S0168-9002(03)01368-8}}.

\bibitem{CMS:2020uim}
\hrefCMSnoop {}{{CMS Collaboration}, ``Electron and photon reconstruction and
  identification with the {CMS} experiment at the {CERN} {LHC}'',} \textit{
  JINST} \textbf{ 16} (2021) P05014,
  \href{http://dx.doi.org/10.1088/1748-0221/16/05/P05014}{\doi{10.1088/1748-0221/16/05/P05014}},
  \href{http://www.arXiv.org/abs/2012.06888}{\texttt{arXiv:2012.06888}}.

\bibitem{CMS:2018rym}
\hrefCMSnoop {}{{CMS Collaboration}, ``Performance of the {CMS} muon detector
  and muon reconstruction with proton-proton collisions at {$\sqrt{s}=$ 13
  TeV}'',} \textit{ JINST} \textbf{ 13} (2018) P06015,
  \href{http://dx.doi.org/10.1088/1748-0221/13/06/P06015}{\doi{10.1088/1748-0221/13/06/P06015}},
  \href{http://www.arXiv.org/abs/1804.04528}{\texttt{arXiv:1804.04528}}.

\bibitem{CMSjetID}
\href {https://cds.cern.ch/record/2256875/}{{CMS Collaboration}, ``{CMS} jet
  algorithms performance in 13 {TeV} data'',} CMS Physics Analysis Summary
  CMS-PAS-JME-16-003, 2017.

\bibitem{ParticleDataGroup:2024cfk}
\hrefCMSnoop {}{{Particle Data Group} Collaboration, ``Review of particle
  physics'',} \textit{ Phys. Rev. D} \textbf{ 110} (2024) 030001,
  \href{http://dx.doi.org/10.1103/PhysRevD.110.030001}{\doi{10.1103/PhysRevD.110.030001}}.

\bibitem{CMS:2021roc}
\hrefCMSnoop {}{{CMS Collaboration}, ``Search for heavy resonances decaying to
  a pair of lorentz-boosted higgs bosons in final states with leptons and a
  bottom quark pair at {$\sqrt{s} = 13~\mathrm{TeV}$}'',} \textit{ JHEP}
  \textbf{ 05} (2022) 005,
  \href{http://dx.doi.org/10.1007/JHEP05(2022)005}{\doi{10.1007/JHEP05(2022)005}},
  \href{http://www.arXiv.org/abs/2112.03161}{\texttt{arXiv:2112.03161}}.

\bibitem{Oreglia:1980cs}
\href
  {https://www.slac.stanford.edu/pubs/slacreports/reports13/slac-r-236.pdf}{M.~J.
  Oreglia, ``A study of the reactions {$\psi^\prime \to \gamma \gamma
  \psi$}''}.
\newblock PhD thesis, Stanford University, 1980.
\newblock {{SLAC} Report {SLAC-R-236}, see Appendix {D}}.

\bibitem{Gaiser:1982yw}
\href
  {http://slac.stanford.edu/pubs/slacreports/reports02/slac-r-255.pdf}{J.~E.
  Gaiser, ``Charmonium spectroscopy from radiative decays of the {$J/\psi$} and
  {$\psi^\prime$}''}.
\newblock PhD thesis, Stanford University, 1982.
\newblock {{SLAC} Report {SLAC-R-255}}.

\bibitem{Punzi:2003bu}
\hrefCMSnoop {}{G.~Punzi, ``Sensitivity of searches for new signals and its
  optimization'',} p.~MODT002.
\newblock 2003.
\newblock
  \href{http://www.arXiv.org/abs/physics/0308063}{\texttt{arXiv:physics/0308063}}.
\newblock {econf C030908}.

\bibitem{Dauncey:2014xga}
\hrefCMSnoop {}{P.~D. Dauncey, M.~Kenzie, N.~Wardle, and G.~J. Davies,
  ``Handling uncertainties in background shapes: the discrete profiling
  method'',} \textit{ JINST} \textbf{ 10} (2015) P04015,
  \href{http://dx.doi.org/10.1088/1748-0221/10/04/P04015}{\doi{10.1088/1748-0221/10/04/P04015}},
  \href{http://www.arXiv.org/abs/1408.6865}{\texttt{arXiv:1408.6865}}.

\bibitem{CMS:2021xjt}
\hrefCMSnoop {}{{CMS Collaboration}, ``Precision luminosity measurement in
  proton-proton collisions at {$\sqrt{s} =$ 13 TeV} in 2015 and 2016 at
  {CMS}'',} \textit{ Eur. Phys. J. C} \textbf{ 81} (2021) 800,
  \href{http://dx.doi.org/10.1140/epjc/s10052-021-09538-2}{\doi{10.1140/epjc/s10052-021-09538-2}},
  \href{http://www.arXiv.org/abs/2104.01927}{\texttt{arXiv:2104.01927}}.

\bibitem{CMS-PAS-LUM-17-004}
\href {https://cds.cern.ch/record/2621960/}{{CMS Collaboration}, ``{CMS}
  luminosity measurement for the 2017 data-taking period at $\sqrt{s}$ = 13
  {TeV}'',} CMS Physics Analysis Summary CMS-PAS-LUM-17-004, 2018.

\bibitem{CMS-PAS-LUM-18-002}
\href {https://cds.cern.ch/record/2676164/}{{CMS Collaboration}, ``{CMS}
  luminosity measurement for the 2018 data-taking period at $\sqrt{s}$ = 13
  {TeV}'',} CMS Physics Analysis Summary CMS-PAS-LUM-18-002, 2019.

\bibitem{CMS:2010vta}
\hrefCMSnoop {}{{CMS Collaboration}, ``Measurement of the {$\PW$} and {$\PZ$}
  inclusive production cross sections at {$ \sqrt{s}=7 $ TeV} with the {CMS}
  experiment at the {LHC}'',} \textit{ JHEP} \textbf{ 10} (2011) 132,
  \href{http://dx.doi.org/10.1007/JHEP10(2011)132}{\doi{10.1007/JHEP10(2011)132}},
  \href{http://www.arXiv.org/abs/1107.4789}{\texttt{arXiv:1107.4789}}.

\bibitem{CMS:2018mlc}
\hrefCMSnoop {}{{CMS Collaboration}, ``Measurement of the inelastic
  proton-proton cross section at {$ \sqrt{s}=13 $} {TeV}'',} \textit{ JHEP}
  \textbf{ 07} (2018) 161,
  \href{http://dx.doi.org/10.1007/JHEP07(2018)161}{\doi{10.1007/JHEP07(2018)161}},
  \href{http://www.arXiv.org/abs/1802.02613}{\texttt{arXiv:1802.02613}}.

\bibitem{CMS:2024ppo}
\hrefCMSnoop {}{{CMS Collaboration}, ``Performance of the {CMS} electromagnetic
  calorimeter in pp collisions at {$\sqrt{s} = 13~\mathrm{TeV}$}'',} \textit{
  JINST} \textbf{ 19} (2024) P09004,
  \href{http://dx.doi.org/10.1088/1748-0221/19/09/P09004}{\doi{10.1088/1748-0221/19/09/P09004}},
  \href{http://www.arXiv.org/abs/2403.15518}{\texttt{arXiv:2403.15518}}.

\bibitem{Butterworth:2015oua}
\hrefCMSnoop {}{J.~Butterworth { et~al.}, ``{PDF4LHC} recommendations for {LHC}
  {R}un {II}'',} \textit{ J. Phys. G} \textbf{ 43} (2016) 023001,
  \href{http://dx.doi.org/10.1088/0954-3899/43/2/023001}{\doi{10.1088/0954-3899/43/2/023001}},
  \href{http://www.arXiv.org/abs/1510.03865}{\texttt{arXiv:1510.03865}}.

\bibitem{Junk:1999kv}
\hrefCMSnoop {}{T.~Junk, ``Confidence level computation for combining searches
  with small statistics'',} \textit{ Nucl. Instrum. Meth. A} \textbf{ 434}
  (1999) 435,
  \href{http://dx.doi.org/10.1016/S0168-9002(99)00498-2}{\doi{10.1016/S0168-9002(99)00498-2}},
  \href{http://www.arXiv.org/abs/hep-ex/9902006}{\texttt{arXiv:hep-ex/9902006}}.

\bibitem{Read:2002hq}
\hrefCMSnoop {}{A.~L. Read, ``Presentation of search results: The {$CL_s$}
  technique'',} \textit{ J. Phys. G} \textbf{ 28} (2002) 2693,
  \href{http://dx.doi.org/10.1088/0954-3899/28/10/313}{\doi{10.1088/0954-3899/28/10/313}}.

\bibitem{Cowan:2010js}
\hrefCMSnoop {}{G.~Cowan, K.~Cranmer, E.~Gross, and O.~Vitells, ``Asymptotic
  formulae for likelihood-based tests of new physics'',} \textit{ Eur. Phys. J.
  C} \textbf{ 71} (2011) 1554,
  \href{http://dx.doi.org/10.1140/epjc/s10052-011-1554-0}{\doi{10.1140/epjc/s10052-011-1554-0}},
  \href{http://www.arXiv.org/abs/1007.1727}{\texttt{arXiv:1007.1727}}.
  [Erratum: \DOI{10.1140/epjc/s10052-013-2501-z}].

\bibitem{Chen:2017hak}
\hrefCMSnoop {}{C.-Y. Chen, S.~Dawson, and E.~Furlan, ``Vectorlike fermions and
  {H}iggs effective field theory revisited'',} \textit{ Phys. Rev. D} \textbf{
  96} (2017) 015006,
  \href{http://dx.doi.org/10.1103/PhysRevD.96.015006}{\doi{10.1103/PhysRevD.96.015006}},
  \href{http://www.arXiv.org/abs/1703.06134}{\texttt{arXiv:1703.06134}}.

\end{thebibliography}\endgroup
\cleardoublepage \appendix\section{The CMS Collaboration \label{app:collab}}\begin{sloppypar}\hyphenpenalty=5000\widowpenalty=500\clubpenalty=5000\cmsinstitute{Yerevan Physics Institute, Yerevan, Armenia}
{\tolerance=6000
A.~Hayrapetyan, V.~Makarenko\cmsorcid{0000-0002-8406-8605}, A.~Tumasyan\cmsAuthorMark{1}\cmsorcid{0009-0000-0684-6742}
\par}
\cmsinstitute{Institut f\"{u}r Hochenergiephysik, Vienna, Austria}
{\tolerance=6000
W.~Adam\cmsorcid{0000-0001-9099-4341}, L.~Benato\cmsorcid{0000-0001-5135-7489}, T.~Bergauer\cmsorcid{0000-0002-5786-0293}, M.~Dragicevic\cmsorcid{0000-0003-1967-6783}, P.S.~Hussain\cmsorcid{0000-0002-4825-5278}, M.~Jeitler\cmsAuthorMark{2}\cmsorcid{0000-0002-5141-9560}, N.~Krammer\cmsorcid{0000-0002-0548-0985}, A.~Li\cmsorcid{0000-0002-4547-116X}, D.~Liko\cmsorcid{0000-0002-3380-473X}, M.~Matthewman, J.~Schieck\cmsAuthorMark{2}\cmsorcid{0000-0002-1058-8093}, R.~Sch\"{o}fbeck\cmsAuthorMark{2}\cmsorcid{0000-0002-2332-8784}, M.~Shooshtari\cmsorcid{0009-0004-8882-4887}, M.~Sonawane\cmsorcid{0000-0003-0510-7010}, W.~Waltenberger\cmsorcid{0000-0002-6215-7228}, C.E.~Wulz\cmsAuthorMark{2}\cmsorcid{0000-0001-9226-5812}
% ...
\par}
$^{1}$Also at Yerevan State University, Yerevan, Armenia\\
$^{2}$Also at Technische Universit\"{a}t Wien, Vienna, Austria\\
% ...
\end{sloppypar}
%%% END EDITABLE REGION %%%
% skeleton_end
\end{document}